\documentclass[12pt]{article}
\pdfoutput=1
\usepackage{epsf,color,colordvi}
\usepackage{graphicx}
\usepackage{amsmath}
\usepackage{amssymb}
\usepackage{epsfig}
\usepackage{cite}
\usepackage{fontenc}
\usepackage{float}
\usepackage[lofdepth,lotdepth,caption=false]{subfig}
\usepackage{color}
\usepackage{booktabs}
\usepackage{tabularx}
\usepackage{multirow}
\usepackage{longtable}
\usepackage{lscape}
\usepackage{dcolumn}
\usepackage{rotating}
\usepackage{hyperref}\usepackage[normalem]{ulem}
\definecolor{darkgreen}{cmyk}{1,0,1,0.4}
\definecolor{brown}{cmyk}{0,0.8,1,0.2}
\definecolor{darkred}{cmyk}{0,1,1,0.2}

\allowdisplaybreaks[4] 
\makeatletter
\renewcommand{\fnum@table}{\textbf{\tablename~\thetable}}
\renewcommand{\fnum@figure}{\textbf{\figurename~\thefigure}}
\makeatother

\newcounter{myenumi}

\renewcommand{\themyenumi}{\roman{myenumi}}
{\end{list}}

\newlength{\myem}
\newcounter{mysubequation}[equation]

\makeatletter
\renewcommand{\section}{\@startsection{section}{1}{0em}{-\baselineskip}%
{\baselineskip}{\normalfont\large\bfseries}}
\renewcommand{\subsection}%
{\@startsection{subsection}{2}{0em}{-0.7\baselineskip}%
{0.7\baselineskip}{\normalfont\bfseries}}
\makeatother

\newcommand{\bi}{\begin{itemize}}
\newcommand{\ei}{\end{itemize}}

\def\beq{\begin{equation}}
\def\eeq{\end{equation}}

\newcommand{\bea}{\begin{eqnarray}}
\newcommand{\eea}{\end{eqnarray}}

\newcommand{\ldm}{\Delta m_{31}^2}

\newcommand{\ie}{{\it i.e.}}

\newcommand{\eet}{\varepsilon_{e\tau}}
\newcommand{\emt}{\varepsilon_{\mu\tau}}
\newcommand{\ett}{\varepsilon_{\tau\tau}}
\newcommand{\eee}{\varepsilon_{ee}}
\newcommand{\eem}{\varepsilon_{e\mu}}
\newcommand{\emm}{\varepsilon_{\mu\mu}}
\newcommand{\eeta}{|\varepsilon_{e\tau}|}

\newcommand{\eema}{|\varepsilon_{e\mu}|}

\newcommand{\eetp}{\varphi_{e\tau}}
\newcommand{\emtp}{\varphi_{\mu\tau}}
\newcommand{\eemp}{\varphi_{e\mu}}
\def\epsilon{\varepsilon}

\newcommand{\ttok}{{\sc T2K}}
\newcommand{\dune}{{\sc DUNE}}

\newcommand{\nova }{{\sc NOvA}}

\newcommand{\ttohk}{{\sc T2HK}}

\newcommand\sch{Schr$\ddot{\rm o}$dinger~}

\def\<{\langle}
\def\>{\rangle}

\def\dfrac#1#2{{\displaystyle\frac{#1}{#2}}}

\def\lsim{\mathrel{\rlap{\lower4pt\hbox{\hskip1pt$\sim$}}
    \raise1pt\hbox{$<$}}}         
\def\gsim{\mathrel{\rlap{\lower4pt\hbox{\hskip1pt$\sim$}}
    \raise1pt\hbox{$>$}}}         

\newcommand{\gev}{\textrm{{GeV}}}
\newcommand{\km}{\textrm{{km}}}

\newcommand{\len}{\textrm{{L}}}
\newcommand{\e}{\textrm{{E}}}

\newcommand{\lbye}{\textrm{{L/E}}}

\newcommand{\sumpcp}[1]{\ensuremath{\sum {P}_{\alpha\beta}  \left(#1 \right)}}

\newcommand{\acp}[1]{\ensuremath{{A}^{CP}_{\alpha\beta} \left(#1 \right)}}

\newcommand{\dacp}[1]{\ensuremath{\delta [\Delta P^{CP/T}_{\alpha\beta}]}}

\newcommand{\pbarab}[1]{\ensuremath{{ P}_{\bar{\alpha} \bar{\beta}} }}
\newcommand{\pbarba}[1]{\ensuremath{{ P}_{\bar{\beta} \bar{\alpha}} }}
\newcommand{\acpab}[1]{\ensuremath{A^{CP}_{\alpha \beta}}}
\newcommand{\acpaa}[1]{\ensuremath{A^{CP}_{\alpha \alpha}}}
\newcommand{\ataa}[1]{\ensuremath{A^{T}_{\alpha \alpha}}}
\newcommand{\acpba}[1]{\ensuremath{A^{CP}_{\beta \alpha}}}
\newcommand{\atab}[1]{\ensuremath{{A}^{T}_{\alpha \beta}}}
\newcommand{\atba}[1]{\ensuremath{{A}^{T}_{\beta \alpha}}}

\newcommand{\acptab}[1]{\ensuremath{A^{CPT}_{\alpha \beta}}}
\newcommand{\dpcpab}[1]{\ensuremath{\Delta {\cal P}^{CP}_{\alpha \beta} }}
\newcommand{\dptab}[1]{\ensuremath{\Delta {\cal P}^{T}_{\alpha \beta}}}
\newcommand{\dpcptab}[1]{\ensuremath{\Delta {\cal P}^{CPT}_{\alpha \beta} }}

\begin{document}

\title{Can we probe intrinsic CP/T violation and non-unitarity at long baseline accelerator experiments ?}
%
\begin{titlepage}

\renewcommand{\thefootnote}{\alph{footnote}}

\vspace*{-3.cm}
\begin{flushright}

\end{flushright}


\renewcommand{\thefootnote}{\fnsymbol{footnote}}
\setcounter{footnote}{-1}

{\begin{center}
{\large\bf 
Can we probe intrinsic CP/T violation and non-unitarity at long baseline accelerator experiments ?%
\\[0.2cm]
}
\end{center}}

\renewcommand{\thefootnote}{\alph{footnote}}

\vspace*{.8cm}
\vspace*{.3cm}
{\begin{center} 
            {{\sf 
                Jogesh Rout$^\S$\,\footnote[1]{\makebox[1.cm]{Email:}
                jogesh.rout1@gmail.com},
                Mehedi Masud${^{\P \,\ast}}$\,\footnote[2]
                {\makebox[1.cm]{Email:} masud@ific.uv.es}
                 and 
                Poonam Mehta$^{\S}$\,\footnote[3]{\makebox[1.cm]{Email:}
                pm@jnu.ac.in}

               }}

\end{center}}
\vspace*{0cm}
{\it 
\begin{center}
$^\S$\, School of Physical Sciences,   \\   Jawaharlal Nehru University, 
      New Delhi 110067, India

$^\P$ \, 
Astroparticle and High Energy Physics Group, Institut de F\'{i}sica Corpuscular --
  C.S.I.C./Universitat de Val\`{e}ncia, Parc Cientific de Paterna.\\
  C/Catedratico Jos\'e Beltr\'an, 2 E-46980 Paterna (Val\`{e}ncia) - Spain

$^\ast$
Harish-Chandra Research Institute,\\ Chhatnag Road, Jhunsi, Allahabad 211 019,
 India   
	
\end{center}}

\vspace*{1.5cm}


{\Large 
\bf
 \begin{center} Abstract  
\end{center} 
 }
 One of the fundamental parameters entering neutrino oscillation framework is the leptonic CP phase $\delta_{13}$ and its measurement is an important goal of the planned long baseline experiments.  It should be noted that ordinary matter effects complicate the determination of this parameter and there are studies in literature that deal with separation of intrinsic versus extrinsic CP violation. It is important to investigate the consequences of new physics effects that can not only hamper the measurement of $\delta_{13}$, but also impact the consequences of discrete symmetries such as CP, T and unitarity in different oscillation  channels. In the present work, we  explore these discrete symmetries and implications on unitarity in presence of two  new physics scenarios (non-standard interaction in propagation and presence of sterile neutrinos) that serve as good examples of going beyond the standard scenario in different directions. We uncover the impact of new physics scenarios on 
 disentangling intrinsic and extrinsic CP violation. 
 
\vspace*{.5cm}

\end{titlepage}

\newpage

\renewcommand{\thefootnote}{\arabic{footnote}}
\setcounter{footnote}{0}

%
\section{Present status of three generation neutrino parameters}
\label{sec:1}

The possibility of neutrino oscillations was first raised in a seminal paper by Pontecorvo~\cite{pontecorvo} 
and almost sixty years later the experimental confirmation of neutrino oscillations was rewarded with a Nobel prize~\cite{nobel2015}. 
The standard three flavour neutrino mixing parameters are - three angles ($\theta_{12},\theta_{13},\theta_{23}$), two mass 
splittings ($\delta m^2_{31}, \delta m^2_{21}$) and one phase ($\delta_{13}$) that is responsible for CP violation in the leptonic sector.
   While the mixing angles and the mass-squared differences (and absolute value of only one of them) have been measured with varying degrees of precision (see Table~\ref{tab:parameters}), the measured value of $\theta_{13}$ allows for an early measurement of the leptonic CP violation~\cite{Capozzi:2013csa,Forero:2014bxa,Esteban:2016qun}.

The three flavour neutrino mixing matrix ${\cal U}$  is parameterized by three angles $\theta_{12},\theta_{23},\theta_{13}$ and one phase  $\delta_{13}$~\footnote{For n flavors, the leptonic mixing matrix 
$U_{\alpha i}$ depends on $(n-1)(n-2)/2$ Dirac-type CP violating  phases. 
If the neutrinos are Majorana particles, there are $(n-1)$ additional, so called Majorana-type CP violating phases.}. In the  Pontecorvo-Maki-Nakagawa-Sakata (PMNS) parametrization~\cite{Beringer:1900zz}, ${\cal U}$ is given by
\bea
{\mathcal U}^{} &=& \left(
\begin{array}{ccc}
c_{12} c_{13}   & s_{12} c_{13} & s_{13} e^{-i\delta_{13}} \\
 -s_{12} c_{23} -c_{12} s_{13} s_{23} e^{i\delta_{13}} &
c_{12} c_{23}  -s_{12} s_{13} s_{23} e^{i\delta_{13}} & c_{13} s_{23}   \\ 
s_{12} s_{23} - c_{12} s_{13} c_{23} e^{i\delta_{13}}  & -c_{12} s_{23} - s_{12} s_{13} c_{23} e^{i\delta_{13}} & c_{13} c_{23} \\
\end{array} 
\right)   
 \ ,
\label{u}
 \eea 
where $s_{ij}=\sin {\theta_{ij}}, c_{ij}=\cos \theta_{ij}$. The validity of the three flavour neutrino paradigm relies very heavily on the assumption of $3 \times 3$ unitarity of the mixing matrix.
 Most of the information about the parameters of the neutrino mixing matrix is gleaned from a vast variety of experiments. We should realize that much of the information originates from $\nu_\mu$ and $\nu_e$ sector via disappearance ($\bar\nu_e\to \bar\nu_e$ in case of reactor experiments and $\bar \nu_\mu \to \bar\nu_\mu$ in case of atmospheric and long baseline experiments) and  appearance ($\nu_\mu \to \nu_e$) measurements in the ongoing and future  long baseline neutrino oscillation experiments. The remaining elements in the mixing matrix are fixed assuming unitarity i.e. probability conservation~\cite{Parke:2015goa}. Clearly, data from neutrino experiments is not sufficient to constrain all the elements of the leptonic mixing matrix~\cite{Farzan:2002ct}. 
 On the other hand, the  assumption of unitarity in the quark sector is well justified by data.

\begin{table}[htb]
\centering
\begin{tabular}{| lccc| }
\hline
&&&\\
Parameter & Best-fit value & 3$\sigma$ range & Precision (\%)  \\
&&&\\
\hline
&&&\\
$\sin^2 \theta_{12}$ & 0.304 & 0.270 $\to$ 0.344 & 12    \\
$\sin^2 \theta_{13}$  & 0.0218   & 0.0186 $\to $ 0.0250  & 14 \\
&0.0219& 0.0188 $\to$ 0.0251 & 14 \\
$\sin^2 \theta_{23}$  & 0.452  & 0.382 $\to$ 0.643  & 25  \\
&0.579& 0.389 $\to$ 0.644&24 \\
&&&\\
$\delta m^2_{21}$ [$10^{-5}$ $eV^2$]  &  7.50  & 7.02 $\to$ 
8.09 &  \\
&&&\\
$\delta m^2_{3l}$  [$10^{-3}$ $eV^2$] & $+2.457$   &  7
$+2.317$ $\to$ $+2.607$ & 6 \\
 & $-2.449$ & $-2.590$ $\to$  $-2.307$ & 6 \\
&&&\\
$\delta_{13}$ & $\ast$ & $[-\pi : \pi]$ & $\ast$\\
&&&\\
\hline
\end{tabular}
\caption{\label{tab:parameters}
The best-fit values, 3$\sigma$ ranges and precision (in percentage) of the six parameters 
from the latest global fit to neutrino data~\cite{Esteban:2016qun}. 
For entries with two rows, the first (second) row corresponds to NH (IH). For NH, 
$\delta m^2_{3l} \equiv \delta m^2 _{31} >0$ and for IH, $\delta m^2_{3l} \equiv \delta m^2 _{32} < 0$.}
\end{table}

Within the Standard Model (SM), the CP symmetry is broken by complex phases 
in the Yukawa couplings. After removing the unphysical phases in the SM,
there is only one physical phase which is the CP violating parameter. This  economical description of CP violation
 in the SM is referred to as the Kobayashi-Maskawa (KM) mechanism~\cite{Kobayashi:1973fv}. 
But if neutrinos are Majorana particles, there can be 
two additional Majorana-type phases in the three flavour case which
 can not be probed via oscillation experiments. 
 In vacuum, the lone CP phase given by KM mechanism accounts for the CP violation signal.
 However standard interaction (SI) along with 
 the inherent CP asymmetry present in the Earth matter  introduces effects that mimic CP violation. 
   These are referred to as fake/extrinsic CP violating effects as opposed to the 
 genuine/intrinsic CP violating effect due to the presence of $\delta_{13}$~\cite{Arafune:1997hd,Bilenky:1997dd,
 BurguetCastell:2001ez,Nunokawa:2007qh,Branco:2011zb,Ohlsson:2013ip}. Earlier attempts have proposed 
 experimental arrangements~\cite{Minakata:2000ee} and observables useful in disentangling the intrinsic and 
 extrinsic CP violation components~\cite{Ohlsson:2013ip} however 
 standard physics was assumed there. Our
  work differs in the fact that we analyse this question in the backdrop of three distinct 
  physics scenarios (SI and two possible sub-dominant new 
   physics effects) and quantitatively demonstrate that new physics further hinders clean extraction of the intrinsic 
   component.    We also quantify our results in terms of event rates and make
    realistic inferences about separability between the intrinsic and extrinsic CP violating effects.   
   
  In the era of precision measurements in neutrino oscillation physics, we need to consider
     subdominant new physics scenarios such as nonstandard neutrino interactions (NSI) or sterile neutrinos
      and discuss the capabilities of our planned  experiments for some bench mark values of new physics parameters. 
We study how neutrino oscillations have the potential to shed light on these crucial questions relating 
CP and T symmetries as well as  the unitarity of the neutrino mixing matrix. We also probe deviations 
due to new physics scenarios. We go beyond the SM in two respects - one in which we introduce 
subdominant effects due to a possible source of new physics dubbed as  NSI~\cite{Masud:2015xva,
deGouvea:2015ndi,Coloma:2015kiu,Masud:2016bvp,Masud:2016gcl,Blennow:2016etl,Forero:2016cmb,
deGouvea:2016pom,Fukasawa:2016lew,Liao:2016orc} and another where the presence of extra sterile
 state can lead to non-unitarity in the $3 \times 3 $ part even though the overall mixing matrix is still 
 unitary~\cite{Gandhi:2015xza,Agarwalla:2016mrc,Agarwalla:2016xxa,Choubey:2016fpi,Dutta:2016glq,Agarwalla:2016xlg,Blennow:2016jkn,Deepthi:2016erc} and examine consequences relevant to long baseline experiments (For other new physics scenarios such as non-unitarity, see~\cite{Ge:2016xya,Dutta:2016vcc,Miranda:2016wdr,Escrihuela:2016ube,Fong:2016yyh}).
We highlight the regions in $L-E$ space where the  effects due to CP and T violation are drastically 
modified due to new physics. Our discussion is mostly targeted towards accelerator based neutrino 
experiments with $L/E \sim 500$ km/GeV but can easily be extended to short baseline experiments and very long baseline experiments.

The plan of the article is as follows. In Section~\ref{cpt}, we give general definitions 
of  CP, T and CPT asymmetries and unitarity condition. In Section~\ref{three}, we give the 
three flavour framework in vacuum (Section~\ref{cptvac}) and in NSI along with the choice 
of NSI parameters (Section~\ref{nsi}). 
In Section~\ref{sterile}, we describe the framework and choice of parameters for the sterile case. 
Our results are discussed in Section~\ref{sec:results}. The CP and T asymmetries are described for the 
   three physics scenarios as a function of $E$ and $L$ in Section~\ref{resa}. For a test of 
   non-unitarity in the sterile case, we use oscillograms as our main tool (see Section~\ref{resb}). 
    In Section~\ref{inex}, we discuss the spectral differences in oscillogram patterns for the three 
    physics scenarios considered~\footnote{For simplicity,  we   assume CP conserving new 
    physics scenarios (\ie, all NSI and sterile phases are set to zero) while obtaining the oscillograms.}. 
    Finally, we discuss prospects of CP violation and implications of 
   our studies for long baseline experiments with particular emphasis 
   on  T2K, T2HK, NOvA and DUNE in Section~\ref{lbl}. We conclude in Section~\ref{con}.

\section{Framework}
\label{sec:framework}

\subsection{CP, T and CPT asymmetries and unitarity condition -  definitions and general remarks} 
\label{cpt}

C, P and T are discrete symmetries that refer to charge conjugation, parity and time
 reversal respectively. 
 CP, T and CPT violation signal in any neutrino oscillation experiment is characterised 
 via a comparison of probability for a given pair of initial and final flavours $\nu_\alpha \to \nu_\beta$
 with their CP, T or CPT conjugate counterparts 
 \bea
 {\textrm{CP  :}} && \quad \nu_{\alpha , \beta} \Longrightarrow \bar\nu_{\alpha ,\beta}  \nonumber\\
  {\textrm{T :}} && \quad \nu_{\alpha,\beta} \Longrightarrow \nu_{\beta,\alpha}  \nonumber\\
   {\textrm{CPT  :}} && \quad \nu_{\alpha,\beta}  \Longrightarrow \bar\nu_{\beta , \alpha}
 \eea
 The action of CP and T are equivalent to complex conjugation of $U_{\alpha i}$.  
 For theoretical discussions on CP, T and CPT in  neutrino oscillations, see \cite{
Akhmedov:2004ve,Akhmedov:2001kd,Xing:2013uxa,Krastev:1988yu,Toshev:1991ku}.
 There can be no CP violation in the two flavour case in vacuum as the unitary matrix in two flavour 
 case can always be made real. In matter with varying density, this need not be the case (for geometric 
 visualization, see~\cite{Mehta:2009ea,Mehta:2009xm}).  Some theoretical consequences of the 
 value of CP violating phase were discussed in~\cite{Farzan:2006vj}. The cosmological effect of 
 CP violation was discussed in~\cite{Khlopov:1981nq}.
 
 Let us define the following asymmetries~\footnote{The denominator $\sumpcp{\delta_{13}}$ has the 
 effect of rescaling the asymmetry curves.} which involve both neutrinos and antineutrinos :
 \begin{eqnarray}
 \label{eq:asymmdef}
     \acpab\ & = & \frac{P_{\alpha \beta} - \bar P_{\alpha \beta}}{P_{\alpha\beta}+\bar 
     P_{\alpha\beta}}=\frac{\Delta P_{\alpha\beta}^{CP}}{\sum P_{\alpha\beta}^{CP}}~, 
     \quad  
    \\
       \atab\  &=& \frac{P_{\alpha \beta} -  P_{\beta \alpha}}{P_{\alpha\beta}+ 
       P_{\beta\alpha}}=\frac{\Delta P_{\alpha\beta}^{T}}{\sum P_{\alpha\beta}^{T}}~, \quad \\
  \acptab\  &=&  \frac{P_{\alpha \beta} - \bar P_{\beta \alpha}}{P_{\alpha\beta}+\bar 
  P_{\beta\alpha}}=\frac{\Delta P_{\alpha\beta}^{CPT}}{\sum P_{\alpha\beta}^{CPT}}~.
 \end{eqnarray}
where $P_{\alpha \beta}$ is the probability for transition $\nu_\alpha \to \nu_\beta$ and 
$\bar P_{\alpha \beta}$ is the probability for transition $\bar\nu_\alpha \to \bar \nu_\beta$. 
The probability expression is given by 

\bea
P_{\alpha\beta} = \sum _{i,j} U_{\alpha i} U_{\beta i}^\star U_{\alpha j}^\star U_{\beta j} \exp\left
\{-i \frac{\delta m^2_{ij}L}{2E}\right\}
\eea
Obviously, these  asymmetries present themselves in different  channels (appearance 
and disappearance) 
	 that can be employed to study CP, T and CPT violation. If CP were exact, the
	  laws of nature would be same for matter and antimatter. While most 
phenomena are CP symmetric, weak interactions violation C and P in the strongest
 possible way. T violation is expected as a corollary of CP violation  if the combined 
 CPT transformation is a fundamental symmetry of Nature.
	 
For three flavour case, there are 9 + 9 appearance and disappearance probability channels for neutrinos and antineutrinos.
Further, the unitarity of the $3\times 3$ mixing matrix (excluding the case of additional sterile neutrino) 
 \begin{equation}
 \sum_{i}  U_{\alpha i} U_{\beta i} ^{\ast}= \delta_{\alpha\beta}~, 
 \end{equation} 
can be translated in terms of probability conservation conditions 
 \bea
 \sum_{\beta} P_{\alpha\beta}  =  \sum_{ \alpha} P_{\alpha\beta} &=& 1~, \\
  \sum_{\beta} \bar P_{\alpha\beta}  = \sum_{\alpha} \bar P_{\alpha\beta} &=& 1~,
 \eea
 which are $6 + 6$ conditions  of which $5 + 5$ are independent. This tells us that $4 (= 9-5) + 4$ 
 neutrino and antineutrino oscillation probabilities are independent.
{{Furthur it may be possible to reduce the number of independent channels to just two for neutrino (and two for anti-neutrino) oscillation probabilities in case of SI. 
In the parameterization considered (Eq.~\ref{u}), the 23 rotation matrix commutes with the matter part in the Hamiltonian. Denoting the $\theta_{23}$ transformed probabilities by
\begin{equation}
\tilde P_{\alpha\beta} \equiv P_{\alpha\beta}
(s_{23}^2 \leftrightarrow c_{23}^2, \sin 2\theta_{23} \to - \sin 2\theta_{23})~, \alpha,\beta=e,\mu,\tau~.
\end{equation}
we can show that
 \begin{equation}
 P_{e \tau} = \tilde P_{e\mu}~,  P_{\tau\mu} = \tilde P_{\mu\tau}~ {\textrm{and}} ~ P_{\tau \tau} = \tilde P_{\mu\mu}~, 
\end{equation} 
and $P_{ee}$ turns out to be independent of $\theta_{23}$. 
Due to unitarity, only two of these are independent~\cite{Akhmedov:2004ve}.
Moreover since the antineutrino probabilities are related to  neutrino probabilities by $\delta_{13} \to -\delta_{13}$ and $A \to -A$, we are left with just two independent
  probabilities (one possible choice could be $P_{e\mu}$ and $P_{\mu\tau}$~\footnote{The choice of these independent probabilities should be such that they should have $\theta_{23}$ dependence and the pair should not be connected by time reversal.}).}}

The unitarity condition  leads to the following  conditions involving CP asymmetries since 
$\sum_{\beta} (P_{\alpha\beta}-\bar P_{\alpha\beta} ) = 0 $ and $\sum_{\alpha} (P_{\alpha\beta}-\bar P_{\alpha\beta} ) = 0 $
 \begin{eqnarray}
  \nonumber \Delta P^{CP}_{e e} + \Delta P^{CP}_{e \mu} + \Delta P^{CP}_{e \tau} = 0~,\\
  \nonumber \Delta P^{CP}_{\mu e} + \Delta P^{CP}_{\mu \mu} + \Delta P^{CP}_{\mu \tau} = 0~,\\
\nonumber \Delta P^{CP}_{\tau e} + \Delta P^{CP}_{\tau \mu} + \Delta P^{CP}_{\tau \tau} = 0~,\\
\nonumber \Delta P^{CP}_{e e} + \Delta P^{CP}_{\mu e} + \Delta P^{CP}_{\tau e} = 0~,\\
  \nonumber \Delta P^{CP}_{e \mu} + \Delta P^{CP}_{\mu \mu} + \Delta P^{CP}_{\tau \mu} = 0~,\\
\Delta P^{CP}_{e\tau} + \Delta P^{CP}_{\mu\tau} + \Delta P^{CP}_{\tau \tau} = 0~.
\label{unitary}
 \end{eqnarray} 
\begin{figure}[htb]
\centering
\includegraphics[width=\textwidth]
{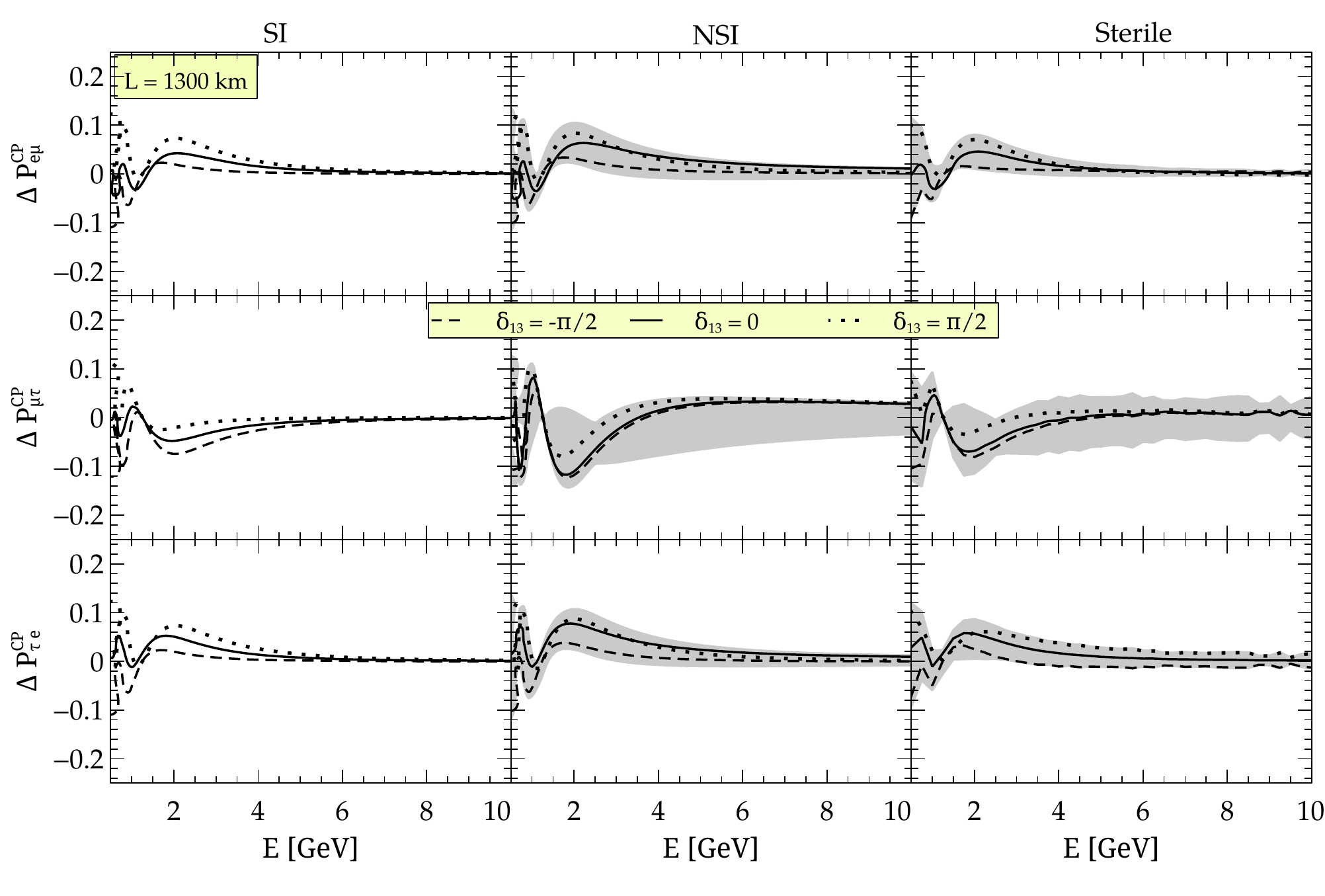}
\caption{\footnotesize{CP odd probability difference ($\Delta P_{\alpha\beta}^{CP}$) plotted as a function of energy at a fixed baseline of $L=1300$ km. The three rows correspond to the different channels considered while the three columns refer to effects due to SI, NSI and Sterile. The solid line corresponds to the case when all CP violating phases are set to zero including $\delta_{13}$. The dotted (dashed) line corresponds to the case when $\delta_{13}=\pi/2 (\delta_{13} = -\pi/2)$ and all additional phases set to zero. The grey bands in the case of NSI and Sterile refer to the variation in phases of the additional parameters introduced in their allowed ranges along with the SI phase $\delta_{13}$ (see Tables~\ref{tab:parameters}, \ref{tab:parameters_nsi} and \ref{tab:parameters_sterile} for the values of the parameters used). 
 }}
\label{fig:1}
\end{figure}
Similarly, in terms of T asymmetries  since $\sum_{\beta} (P_{\alpha\beta}- P_{\beta \alpha} ) = 0 $ and  $\sum_{\alpha} (P_{\alpha\beta}- P_{\beta \alpha} ) = 0 $
 \begin{eqnarray}
  \nonumber \Delta P^{T}_{e e} + \Delta P^{T}_{e \mu} + \Delta P^{T}_{e \tau} = 0\\
  \nonumber \Delta P^{T}_{\mu e} + \Delta P^{T}_{\mu \mu} + \Delta P^{T}_{\mu \tau} = 0\\
 \nonumber \Delta P^{T}_{\tau e} + \Delta P^{T}_{\tau \mu} + \Delta P^{T}_{\tau \tau} = 0\\
  \nonumber \Delta P^{T}_{e e} + \Delta P^{T}_{ \mu e} + \Delta P^{T}_{ \tau e} = 0\\
  \nonumber \Delta P^{T}_{e \mu } + \Delta P^{T}_{\mu \mu} + \Delta P^{T}_{\tau \mu} = 0\\
 \Delta P^{T}_{e \tau} + \Delta P^{T}_{\mu \tau} + \Delta P^{T}_{\tau \tau} = 0
 \end{eqnarray}

\begin{figure}[htb]
\centering
\includegraphics[width=\textwidth]
{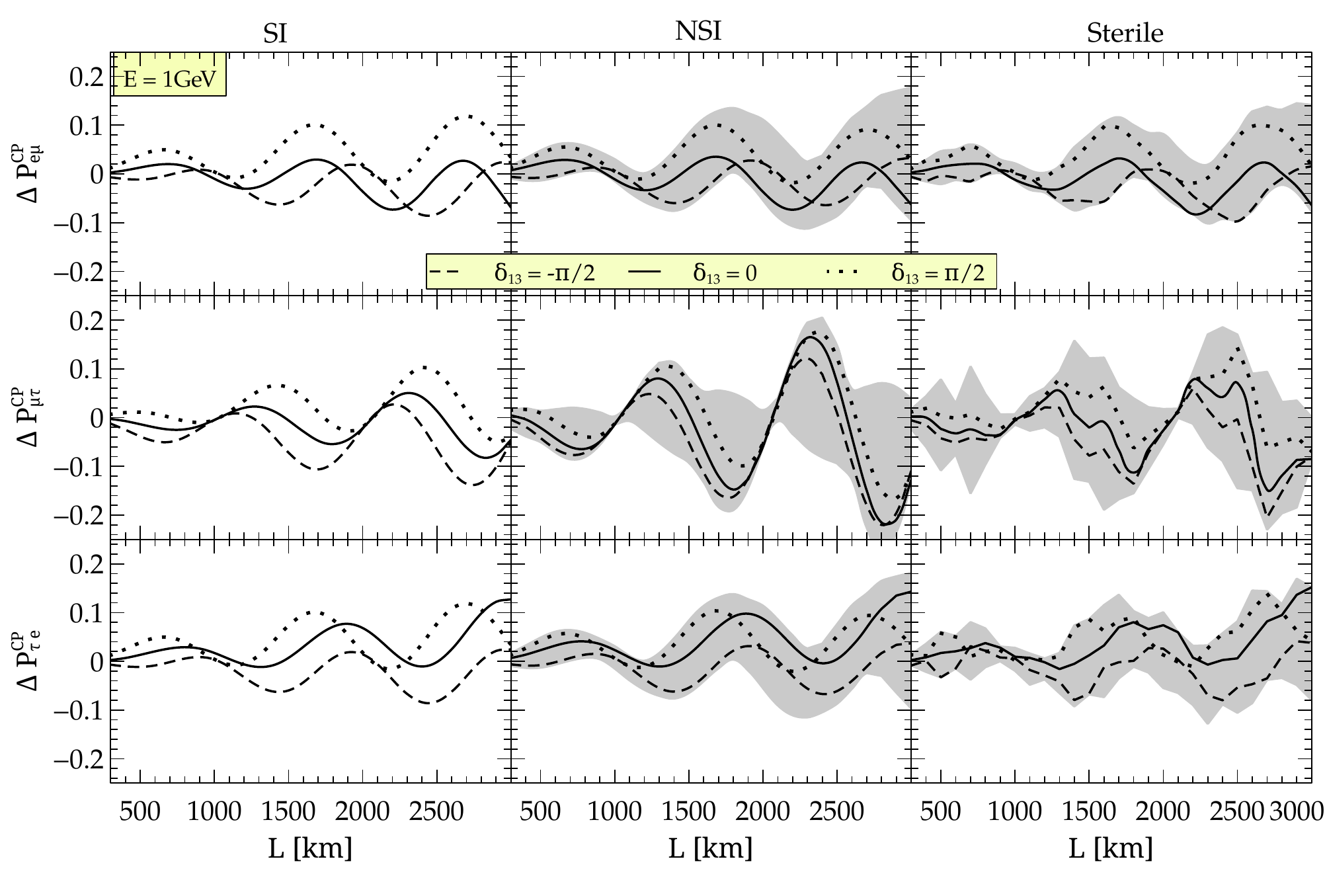}
\caption{\footnotesize{
CP odd probability difference ($\Delta P_{\alpha\beta}^{CP}$) plotted as a function of 
baseline at a fixed energy of $E=1$ GeV. The three rows correspond to the different 
channels considered while the three columns refer to effects due to SI, NSI and Sterile. 
The solid line corresponds to the case when all CP violating phases are set to zero 
including $\delta_{13}$. The dotted (dashed) line corresponds to the case when
 $\delta_{13}=\pi/2 (\delta_{13} = -\pi/2)$ and all additional phases set to zero. 
 The grey bands in the case of NSI and Sterile refer to the variation in phases 
 of the additional parameters introduced in their allowed ranges along with the 
 SI phase $\delta_{13}$ (see Tables~\ref{tab:parameters}, \ref{tab:parameters_nsi} 
 and \ref{tab:parameters_sterile} for the values of the parameters used). 
 }}
\label{fig:2}
\end{figure}

\begin{figure}[htb]
\centering
\includegraphics[width=\textwidth]
{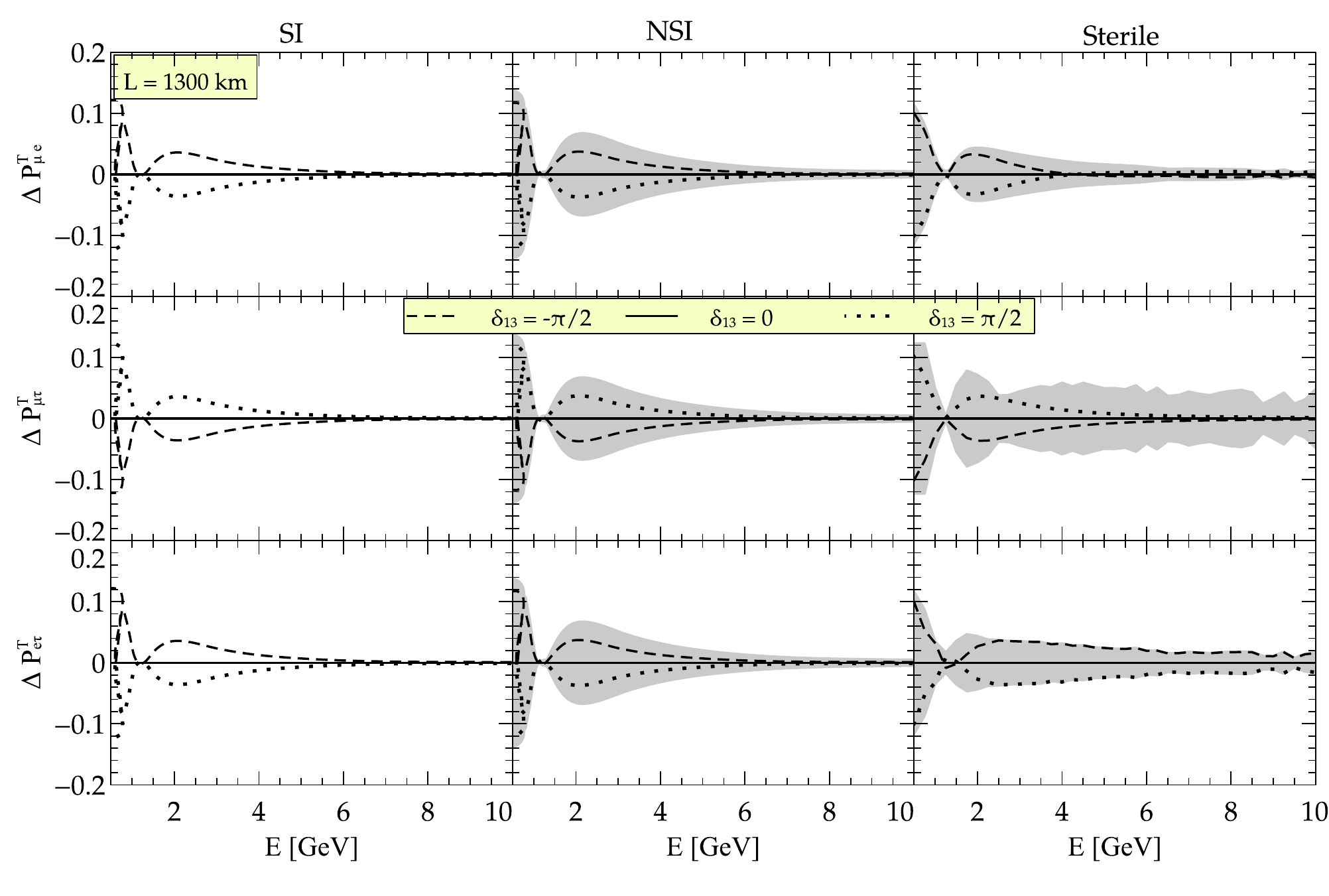}
\caption{\footnotesize{Same as Fig.~\ref{fig:1} but for T asymmetry. %
 }}
\label{fig:3}
\end{figure}

\begin{figure}[htb]
\centering
\includegraphics[width=\textwidth]
{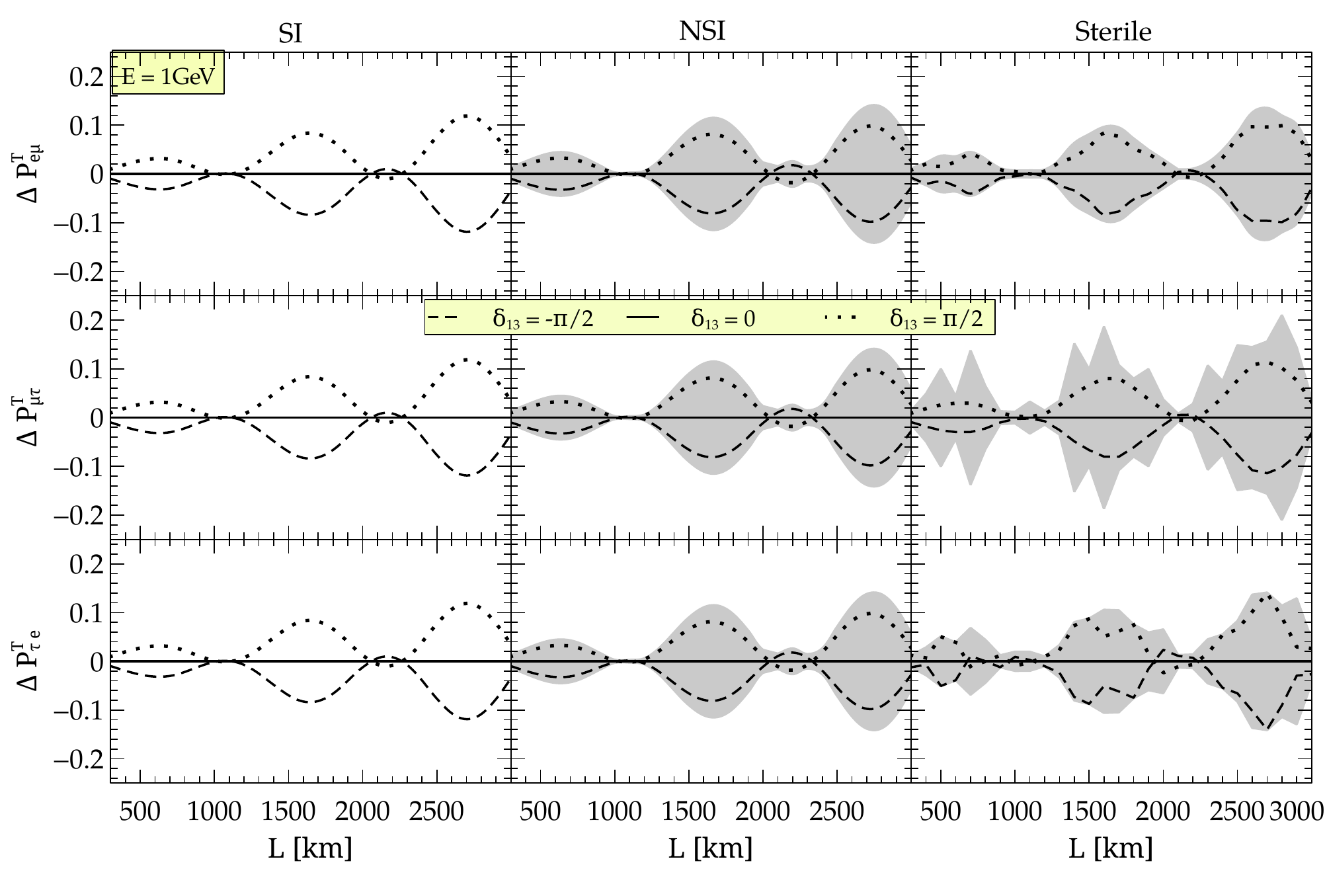}
\caption{\footnotesize{Same as Fig.~\ref{fig:2} but for T asymmetry. %
 }}
\label{fig:4}
\end{figure}

\subsection{Three active neutrinos only}
\label{three}

As is well-known, the three neutrino flavor states can be mapped to a 
three-level quantum system with distinct energy eigenvalues, 
$E_i = p + m^2_i /2p$ in the ultrarelativistic limit in vacuum along 
with the assumption of equal fixed momenta (or energy). 
In the presence of matter, the relativistic dispersion relation $E_i = f(p,m_i)$ gets modified due 
to the neutrino matter interactions during propagation. 

\subsubsection{Propagation in vacuum} 
\label{cptvac}

The effective Hamiltonian is given by 
 \bea
 \label{hexpand_vac} 
 {\mathcal
H}^{}_{\mathrm{v}} &
=&\lambda \left\{ 
{\mathcal U} \left(
\begin{array}{ccc}
0   &  &  \\  &  r_\lambda &   \\ 
 &  & 1 \\
\end{array} 
\right) 
{\mathcal U}^\dagger  
 \right\}  \ ,
 \eea 
where
\begin{equation}
\lambda \equiv \frac{\delta m^2_{31}}{2 E}  \quad \quad ; \quad \quad
r_{\lambda} \equiv \frac{\delta m^2_{21}}{\delta m^2_{31}} \quad \quad  \ .
\label{dimless_vac}
\end{equation}

We first briefly review the case of CPT conservation (CPTC)  \ie,  $\acptab\  = 0$ which  
immediately relates CP and T asymmetries  as 
 \begin{eqnarray}
   \acpab\ = -\acpba\ =  \atab\ = - \atba\ ~ \quad &{\textrm{and}}& \quad   \acpaa\ = 0 = \ataa~.
 \end{eqnarray}
Due to this, the CP asymmetry vanishes when $\alpha = \beta$ (disappearance channels). Further, if 
we assume unitarity of the mixing matrix,  the CP and T asymmetries in the appearance channels are equal to one another in vacuum 
\begin{eqnarray}
\label{eq:cpt_conserved_unitarity}
&&
A^{CP}_{e\mu} = A^{CP}_{\mu\tau} = A^{CP}_{\tau e} = A^{T}_{e\mu} = A^{T}_{\mu\tau} = A^{T}_{\tau e}\nonumber\\ 
&\Rightarrow& \Delta P^{CP}_{\alpha\beta} = \Delta P^{T}_{\alpha\beta} =  \Delta P~,
\end{eqnarray}
where the probability difference in the appearance channels responsible for CP (T) violation is given by $\Delta P$. 
 Now in vacuum, we showed that Eq.~\ref{eq:cpt_conserved_unitarity} holds which means that the CP and T
  violating probability differences in appearance channels are equal to one another. 
  The numerator in the  CP asymmetry (defined in Eq.~\ref{eq:asymmdef}) is given by 
\begin{eqnarray} 
\label{eq:a1_cp}
\Delta P^{CP}_{\alpha\beta}  &=&  8 \, {J} \,\left[
 \sin (r_\lambda \lambda L) \sin^2 \frac{\lambda L}{2} - \sin (\lambda L) \sin^2 \frac{r_\lambda \lambda L}{2} \right] \nonumber \\ 
&=&   4 \, \sin\delta_{13} \, {J}_r\, \left[ \sin\lambda L/2 \sin r_\lambda \lambda L/2 \sin {(1-r_\lambda) \lambda L/2} \right]~,
\end{eqnarray}
where   $J = s_{12} c_{12} s_{23} c_{23} s_{13} c_{13}^2 \sin \delta_{13}$ is the Jarlskog invariant and
 $J_r=J/\sin \delta_{13}$. The second line is obtained after rearranging the terms in the first line. 
 In order to have observable effects, we should have sizeable interference terms that involve the CP violating 
 phase $\delta$. This implies that both $\lambda L/2$ as well as  $r_\lambda \lambda L/2$ must be taken 
 into account. Naturally,  the $\acp{\delta_{13}}$ vanishes as $\delta_{13} \to 0,\pi$ and when 
 $\delta_{13} = \pm \pi/2$,  $\acp{\delta_{13}}$ attains maximal values.
Also it can be noted that the normalised $\acp{\delta_{13}}$ grows linearly with $L/E$.  
  CPT invariance implies CP violation implies T violation and vice versa.

Since the three CP/T odd probability differences (involving different channels) are equal to one another, 
it suffices to look for CP violation in only one of the three possible channels (say $\nu_\mu \to \nu_e$) 
and the conclusions drawn (i.e. whether CP/T is conserved or violated) can be safely extended to 
include the remaining channels which may be difficult to explore otherwise (for example, 
$\nu_\mu \to \nu_\tau$ etc).  Therefore CP conservation or violation in case of vacuum can be 
established by looking at any one of the 
 asymmetries.

\subsubsection{Propagation in matter with effects due to NSI switched on} 
\label{nsi}

\begin{table}[ht]
\centering
\begin{tabular}{ |l l l l l l lll|}
\hline
&&&&&&&&\\
\quad$\varepsilon_{ee}$ \quad & \quad $|\varepsilon_{e\mu}|$\quad & \quad$|\varepsilon_{e\tau}|$ \quad & \quad$|\varepsilon_{\mu e}|$ \quad & \quad $\varepsilon_{\mu\mu}$ \quad & \quad $|\varepsilon_{\mu\tau}|$ \quad & \quad $|\varepsilon_{\tau e}|$ \quad & \quad $|\varepsilon_{\tau\mu}|$ \quad & \quad $\varepsilon_{\tau\tau}$\quad \\
&&&&&&&&\\
\hline
&&&&&&&&\\
 $<$ 4.2 \quad & \quad $<$ 0.3 \quad & \quad $<$ 3.0 \quad & \quad $<$ 0.3 \quad & \quad  $\ast$\quad &\quad $<$ 0.04 \quad &\quad $<$ 3.0 \quad & \quad $<$ 0.04 \quad & \quad $<$ 0.15  \\
&&&&&&&&\\
\hline
\end{tabular}
\caption{\label{tab:parameters_nsi}
The current bounds on NSI parameters taken from~\cite{Biggio:2009nt} (see 
also  \cite{Davidson:2003ha,Gonzalez-Garcia:2013usa}). 
The  phases $\varphi_{\alpha\beta}$ of the off-diagonal parameters are unconstrained 
and can lie the allowed range, $\varphi_{\alpha\beta} \in [-\pi,\pi]$.  }
\end{table}

 The effective Hamiltonian in flavour basis entering the 
 \sch equation for neutrino propagation is given by
 \bea
 \label{hexpand} 
 {\mathcal
H}^{}_{\mathrm{f}} &=&   {\mathcal
H}^{}_{\mathrm{v} } +  {\mathcal
H}^{}_{\mathrm{SI} } +  {\mathcal
H}^{}_{\mathrm{NSI}} 
\nonumber 
\\
&
=&\lambda \left\{ {\mathcal U} \left(
\begin{array}{ccc}
0   &  &  \\  &  r_\lambda &   \\ 
 &  & 1 \\
\end{array} 
\right) {\mathcal U}^\dagger  + r_A   \left(
\begin{array}{ccc}
1  & 0 & 0 \\
0 &  0 & 0  \\ 
0 & 0 & 0 \\ 
\end{array} 
\right)  +
 {r_A}   \left(
\begin{array}{ccc}
\epsilon_{ee}  & \epsilon_{e \mu}  & 
\epsilon_{e \tau}  \\ {\epsilon_{e\mu} }^ \star & 
\epsilon_{\mu \mu} &   \epsilon_{\mu \tau} \\ 
{\epsilon_{e \tau}}^\star & {\epsilon_{\mu \tau}}^\star 
& \epsilon_{\tau \tau}\\
\end{array} 
\right) \right\}  \ ,
 \eea 
where 
\begin{equation}
 r_{A} \equiv \frac{A (x)}{\delta m^2_{31}} \ .
\label{dimless}
\end{equation}
and  
 $A (x)= 2 \sqrt{2}  	E G_F n_e (x)$  where  $n_e$ is the electron
number density. 

The three terms in Eq.~\ref{hexpand} are due to vacuum, matter with 
SI and matter with NSI respectively. For the NSI case,  the 
${\varepsilon}_{\alpha \beta} \, (\equiv |\varepsilon _
{\alpha \beta}|\, e^{i \varphi_{\alpha\beta}})$ are complex  
parameters which appear in ${\cal H}_{NSI}$. 
As a result of the hermiticity of the Hamiltonian, we have nine additional 
parameters (three phases and six amplitudes appearing ${\cal H}_{NSI}$).
 
Switching on matter effects destroys nice feature of the equality of CP and T asymmetries in case of vacuum (see Eq.~15) due to the fact that matter effects  can  fake  the CP violation signal. In such a scenario, establishing CP conservation/violation in a particular channel is not enough to conclude overall CP/T violation in the leptonic sector in general and one needs to examine asymmetries in all possible channels. Even if it the case, one needs to make sure whether the source of CP violation is genuine or fake (depending on the baseline)~\cite{Xing:2000gg}.  In presence of NSI, this is further disturbed due to the presence of additional parameters (see~\cite{Masud:2015xva,Masud:2016bvp,Masud:2016gcl}). On top of that in case of NSI, there are new genuine sources of CP violation as well as new fake sources of CP violation (aka matter effects) that can change the asymmetries even further. For models with possibilties of large NSI, see~\cite{Antusch:2008tz,Farzan:2015doa,Farzan:2015hkd,Forero:2016ghr}.

\subsection{Three active and one sterile neutrino} 
\label{sterile}

This case corresponds to a four level quantum system with the $4 \times 4$ unitary matrix given by 
\bea
{\mathcal U}^{sterile} &=& {\mathcal O}_{34} {\mathcal O}_{24}{\mathcal O}_{14}{\mathcal O}_{23}{\mathcal O}_{13}{\mathcal O}_{12}\nonumber\\
{\mathcal O}_{24}
&=&
\left(
\begin{array}{cccc}
1   & 0 & 0 & 0 \\ 0 & c_{24} & 0 &  e^{-i\delta_{24}} s_{24}   \\ 
0 & 0 & 1 & 0 \\
0 & -e^{i\delta_{24}} c_{24}&0& c_{24}\\
\end{array} 
\right)   
 \ ; {\mathcal O}_{14}
=
\left(
\begin{array}{cccc}
c_{14}   & 0 & 0 & s_{14} \\ 0 & 1 & 0 &  0   \\ 
0 & 0 & 1 & 0 \\
-s_{14} & 0&0& c_{14}\\
\end{array} 
\right)   
 \ , 
\label{us}
 \eea 
 In general ${\mathcal O}_{ij}$ is a rotation matrix in $i-j$ plane parameterized by angle $\theta_{ij}$ and phase $\delta_{ij}$.

In presence of an additional sterile neutrino, the conditions 
 \begin{equation}
 \sum_{\beta}  P_{\alpha\beta} = 1 = \sum_{\beta} \bar P_{\alpha\beta} 
 \end{equation}
 are valid if $\beta$ takes values $e,\mu,\tau,s$ ("s" stands for sterile).  
 The unitarity condition  leads to the following  conditions involving CP asymmetries since $\sum_{\beta} (P_{\alpha\beta}-\bar P_{\alpha\beta} ) = 0 $ 
 \begin{eqnarray}
  \nonumber  \sum_{\alpha=e,\mu,\tau} \Delta P^{CP}_{e \alpha} + \Delta P^{CP}_{e s} = 0\\
  \nonumber  \sum_{\alpha=e,\mu,\tau} \Delta P^{CP}_{\mu \alpha}  + \Delta P^{CP}_{\mu s} = 0\\
\nonumber    \sum_{\alpha=e,\mu,\tau} \Delta P^{CP}_{\tau \alpha} +\Delta P ^{CP}_{\tau s}= 0\\
 \sum_{\alpha=e,\mu,\tau} \Delta P^{CP}_{s \alpha} +\Delta P ^{CP}_{s s}= 0
 \end{eqnarray}
 and $\sum_{\alpha} (P_{\alpha\beta}-\bar P_{\alpha\beta} ) = 0 $ would give four more conditions. 
Similarly, we can get conditions in terms of T asymmetries as $\sum_{\beta} (P_{\alpha\beta}- P_{\beta \alpha} ) = 0 $ and  $\sum_{\alpha} (P_{\alpha\beta}- P_{\beta \alpha} ) = 0 $. Since we can only detect the active flavours, the presence of sterile neutrinos can be felt via flavour dependent measure of non-unitarity.

The choice of parameters used is given in Table~\ref{tab:parameters_sterile}.
\begin{table}[ht]
\centering
\begin{tabular}{ |l l l l l ll|}
\hline
&&&&&&\\
\quad$\theta_{14}$ [$^\circ$] \quad & \quad $\theta_{24}$ [$^\circ$]\quad & \quad$\theta_{34}$ [$^\circ$] \quad & \quad$ \delta_{13}$ [$^\circ$] \quad & \quad $\delta_{24}$ [$^\circ$] \quad & \quad $\delta_{34}$ [$^\circ$] \quad & \quad $\delta m^2_{41} $ [eV$^2$] \quad  \\
&&&&&&\\
\hline
&&&&&&\\
  8.0 \quad & \quad 5.0 \quad & \quad 15.0 \quad & \quad $[-\pi,\pi]$  \quad & \quad  $[-\pi,\pi]$\quad &\quad $[-\pi,\pi]$ \quad &\quad 1.0    \\
&&&&&&\\
\hline
\end{tabular}
\caption{\label{tab:parameters_sterile}
The current bounds on sterile parameters taken from~\cite{An:2014bik,jones_icecube2016,Adamson:2011ku,Kopp:2013vaa,Collin:2016aqd}. 
 }
\end{table}


\begin{figure}[htb]
\centering
\includegraphics[width=1.0\textwidth] 
{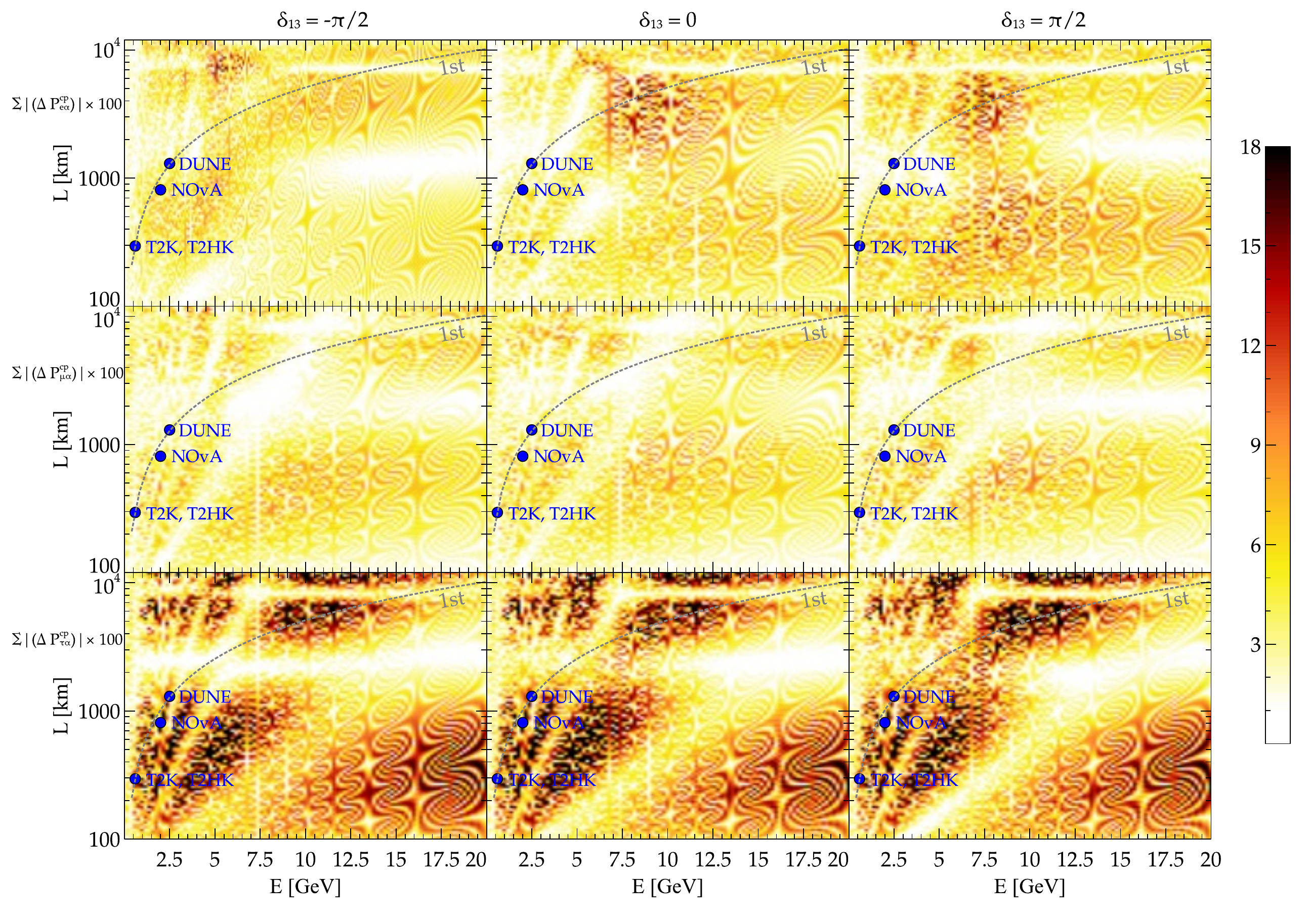}
\caption{\footnotesize{
Measure of non-unitarity ($\sum_{\alpha} |\Delta P^{CP}_{e\alpha}|$, $\sum_{\alpha} |
\Delta P^{CP}_{\mu\alpha} |$, $\sum_{\alpha} |\Delta P^{CP}_{\tau\alpha}|$) in sterile neutrino 
case shown in the plane of $E-L$. The additional phases $\delta_{24},\delta_{34}$ are set to zero.   The location of first oscillation maximum  (for $P_{\mu e}$) is depicted by dashed grey curve. Darker regions imply larger amount of non-unitarity present (in percentage) for those values of $E$ and $L$.}}
\label{fig:9m}
\end{figure}

\begin{figure}[htb]
\centering
\includegraphics[width=1.0\textwidth] 
{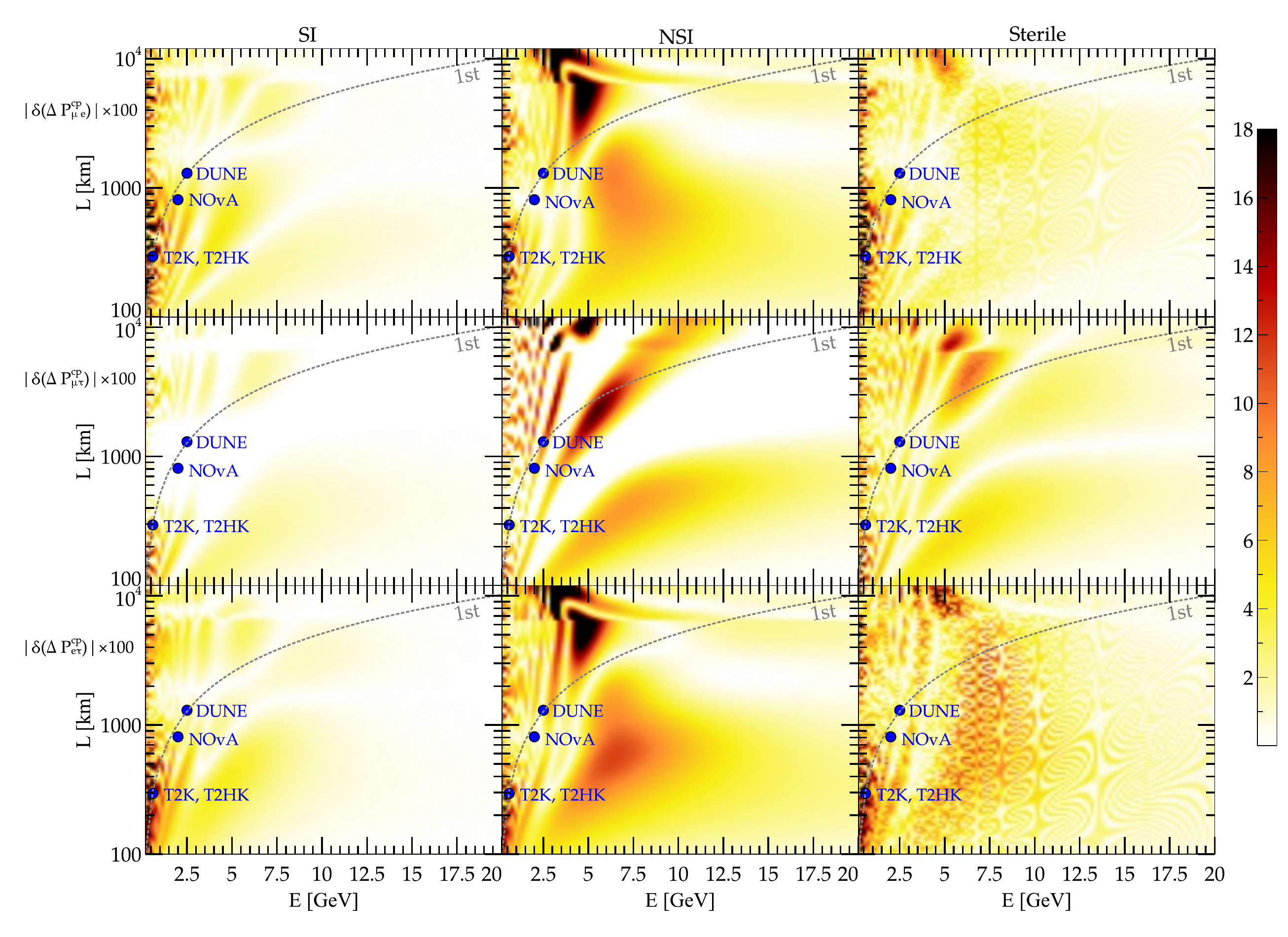}
\caption{\footnotesize{
Oscillogram of absolute relative CP asymmetry for the appearance channels. The location of first oscillation maximum  (for $P_{\mu e}$) is depicted by dashed grey curve. Darker regions imply larger amount absolute relative CP asymmetry (in percentage) for those values of $E$ and $L$.}}
\label{fig:9}
\end{figure}

\begin{figure}[htb]
\centering
\includegraphics[width=1.0\textwidth] 
{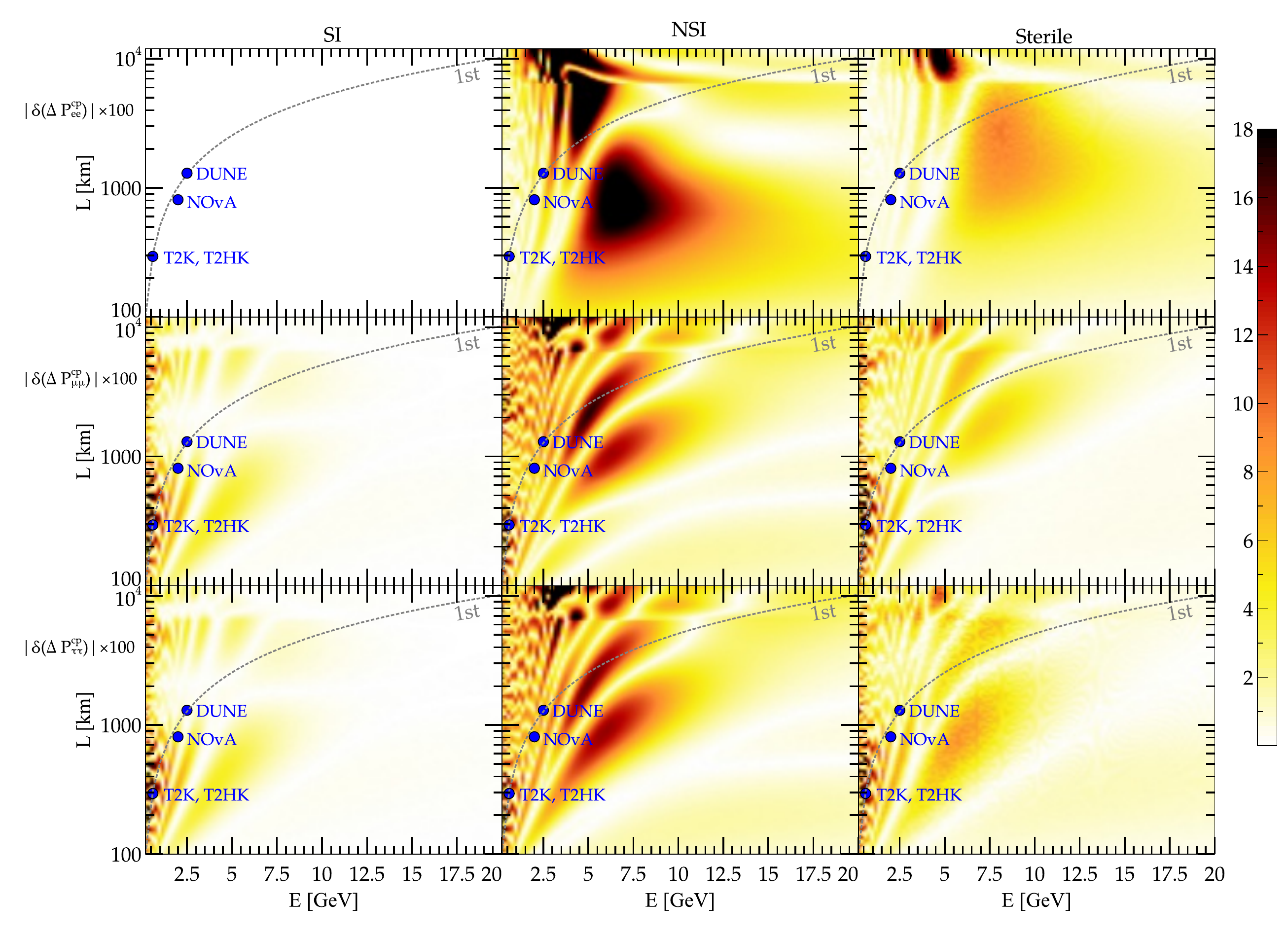}
\caption{\footnotesize{
Oscillogram of absolute relative CP asymmetry for the disappearance channels.  The location of first oscillation maximum  (for $P_{\mu e}$) is depicted by dashed grey curve. Darker regions imply larger amount absolute relative CP asymmetry (in percentage) for those values of $E$ and $L$.}}
\label{fig:10}
\end{figure}

\begin{figure}[htb]
\centering
\includegraphics[width=1.0\textwidth] 
{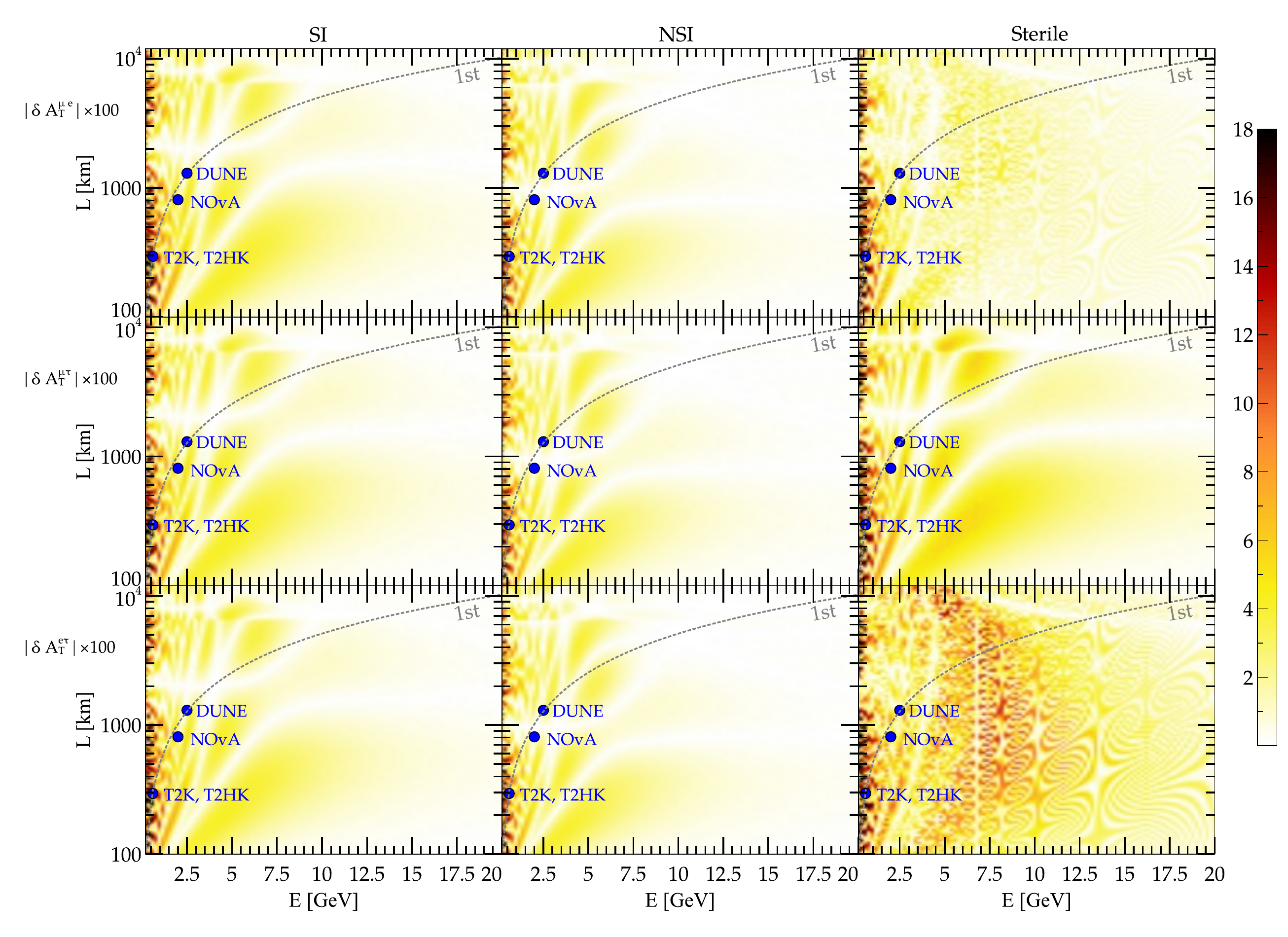}
\caption{\footnotesize{
Oscillogram of absolute relative T asymmetry for the appearance channels. The location of first oscillation maximum  (for $P_{\mu e}$) is depicted by dashed grey curve. Darker regions imply larger amount absolute relative T asymmetry (in percentage) for those values of $E$ and $L$.} }
\label{fig:11}
\end{figure}

\section{Results and Discussion}
\label{sec:results}

\subsection{CP and T asymmetries as a function of $E$ and $L$}
\label{resa}
The CP asymmetries in three appearance channels are plotted as a function of $\e$~and $\len$~ in Fig.~\ref{fig:1} and \ref{fig:2} for fixed baseline of $\len=1300 \km$ and fixed energy of $\e=1~\gev$ respectively. Also, the T asymmetries are plotted in   Fig.~\ref{fig:3} and \ref{fig:4}.
 In all the figures, the three rows correspond to the three different channels while the three columns correspond to the case of SI, NSI and sterile neutrinos. The solid line refers to the case when all CP violating phases including $\delta_{13}$ are set to zero. The dotted (dashed) lines refer to $\delta_{13} = \pi/2 (\delta_{13}=-\pi/2)$ and all additional phases set to zero. The grey bands correspond to the variation in phases of additional parameters in presence of new physics (see previous section). We can infer the following from these plots. 
 
 \begin{itemize}
 \item 
{\sl{CP  asymmetry :}} One would expect the CP asymmetry to vanish when $\delta_{13}=0$ in vacuum (see Eq.~\ref{eq:a1_cp}). From Fig.~\ref{fig:1} and \ref{fig:2}, we note that in all the three physics scenarios, we get non-trivial effects. The size of the effect depends upon the channel and the choice of parameters considered. For example, for a fixed baseline of 1300 km as considered in Fig.~\ref{fig:1} and ${\nu_\mu \to \nu_e}$ channel, we notice that for $\delta_{13}=0$, there is a non-trivial effect due to matter even in the case of SI which is prominent near lower energies (around 4-8\% near the peak). The magnitude is similar in case of NSI and Sterile near the peak.  However there are spectral differences in the three cases which may or may not be visible depending on the particular channel. 
 
 In Fig.~\ref{fig:2}, we plot the CP asymmetry as a function of $len$~ for a fixed value of $\e=1$ \gev. The three curves (solid, dashed and dotted) are oscillatory in nature and we note that there exist specific values of baselines for which one cannot distinguish the curves for the cases $\delta_{13}=\pm\pi/2$ and $\delta_{13}=0$. These lie near $1000$ $\km$~ and $2000$ $\km$. In $\Delta P^{CP}_{\mu\tau}$, we note that surprisingly all the three curves meet near these values of  baseline.
 
 Here again for $\delta_{13}=0$, there is non-trivial effect due to Earth matter and the size of the effects increases  with  baseline and prominently for the new physics scenarios. The spectral differences are also visible here. 
 
 \item  {\sl{T asymmetry :}} We note that $\delta_{13}=0$ (solid line in Fig.~\ref{fig:3}) leads to vanishing asymmetry in all the three cases (SI, NSI and sterile). Also, the dotted and dashed lines are similar in all the three physics cases except for the $e - \tau$ channel in sterile neutrino case. Of course, if additional phases are present, there are non-trivial spectral differences which can be seen as grey bands. 
  
   In Fig.~\ref{fig:4}, we plot the T asymmetry as a function of $\len$~ for a fixed value of $\e=1$ \gev. The three curves (solid, dashed and dotted) are oscillatory in nature and we note that at specific values  (or range of values) of baselines, the three cases ($\delta_{13}=\pm\pi/2$, $\delta_{13}=0$) are indistinguishable. The spectral differences are also visible more prominently in the sterile case. 
  
  \end{itemize}
 
 \subsection{Test of unitarity violation - oscillograms}
\label{resb}

We use coloured oscillograms in the plane of $\e$~and $\len$~ as our tool to depict our observations. 
For the case of SI and NC NSI, the three flavour unitarity is maintained and therefore if we plot the sum of CP odd probability differences, we expect to get blank regions in these cases. 
However, for the case of one additional sterile neutrino, we obtain what is shown in Fig.~\ref{fig:9m}.  
As we can see, there is pattern appearing in the plot and this has been explained  in Appendix~\ref{app_a}. {{Darker regions imply larger amount of non-unitarity present (in percentage) for those values of $\e$~ and $\len$~ as depicted by the colour bar on the right.}}
Primarily, the wiggles are arising due to the large $\delta m^2$ oscillations in the $1-4$ sector. 
 For long baseline neutrino experiments, $\sin^2 (\lambda L/2) \simeq {\cal O} (1)$ which gives 
\begin{eqnarray}
\frac{\lambda L}{2} &\simeq& 1.57~ \left[ \frac{\delta m^2_{31}}{2.5 \times 10^{-3}~eV^2} \frac{2.5~GeV}{E} \frac{L}{1300~km} \right]     ~ \quad \quad {\rm{for ~\dune}}, 
\end{eqnarray}
for the first oscillation maximum (minimum) in the appearance (disappearance) channel. We note that $\e=1.5~\gev, \len=810~\km$ for \nova \, and $\e=0.6~\gev, \len
=295~\km$ for \ttok\, (and also \ttohk) also lead to $\lambda L \sim \pi$.
The location of first oscillation maximum  (using $P_{\mu e}$) is also shown on the plot. Obviously, most of the ongoing or planned long baseline experiments lie on or close to\,\footnote{NOvA is an off-axis experiment.} this curve. 

The different colours depict the amount of unitarity violation. We can note that for the $\nu_e \to \nu_\alpha$ and the $\nu_\tau \to \nu_\alpha$ channel (where $\alpha=e,\mu,\tau$), the violation of unitarity is  larger compared to the $\nu_\mu \to \nu_\alpha$ channel. 
This feature can be ascribed to the larger values of  $\theta_{14} $ and $ \theta_{34}$ used in comparison to $\theta_{24}$ (see Table~\ref{tab:parameters_sterile}). There is also a mild dependence on the value of $\delta_{13}$ as can be seen from different columns. 
Also, if we look at middle row ($\nu_\mu \to \nu_\alpha$ channel), it seems that none of the ongoing long baseline experiments 
can detect the presence of non-unitarity better than $\sim 6\%$ or so.

\subsection{Distinguishing intrinsic and extrinsic CP effects}
\label{inex}

In the context of long baseline experiments where matter can induce extrinsic CP effects, a non-zero value of $\Delta P^{CP}_{\alpha\beta}$ does not unequivocally imply intrinsic CP violation arising due to the Dirac CP phase ($\delta_{13}$). To get over the problem of finding the source of CP violation (i.e. whether 
due to intrinsic CP phase and due to the matter effects), other observables have been introduced~\cite{Ohlsson:2013ip} which can prove useful not only to establish whether CP violation effects arise purely due to the Dirac type CP phase ($\delta_{13}$) or a combination of the intrinsic and extrinsic CP effects. We can define an observable quantity which is obtained by taking difference of the CP probability differences computed at different values of $\delta_{13}$ for the appearance as well as disappearance channels as follows~\cite{Arafune:1997hd,Nunokawa:2007qh,Ohlsson:2013ip} 
\bea
\delta [\Delta P^{CP}_{\alpha\beta}]  &=& [\Delta P_{\alpha\beta}^{CP}] (\delta_{13} = \pi/2) - [\Delta P_{
\alpha\beta}^{CP}] (\delta_{13}=0)~,\nonumber\\
&=& [P_{\alpha\beta} - \bar P _{\alpha\beta}] (\delta_{13} = \pi/2) - 
[P_{\alpha\beta} - \bar P _{\alpha\beta}] (\delta_{13}=0)~.
\label{eq:asymmdif}
\eea
The choices of $\delta_{13}$ in the above equation  allow for better observability of the intrinsic CP effects. 
 Obviously, in vacuum, the second term on the RHS vanishes and the first term gives completely intrinsic CP contribution which will be non-zero for $\delta_{13}=\pi/2$. 
 If matter effects are switched off, Eq.~\ref{eq:asymmdif} reduces to the expression for vacuum asymmetry corresponding to CP violation (see Eqs.~\ref{eq:cpt_conserved_unitarity} and \ref{eq:a1_cp}). 
 In standard matter, both first and second terms on RHS will be non-zero in general. The second term being non-zero in matter signals the presence of purely extrinsic effects. Under certain conditions\,\footnote{As we shall see, both vacuum and matter effects lead to same difference in probability differences due to interesting conspiracy near the first peak~\cite{Marciano:2006uc}. 
  Note that this cancellation is perfect in case of standard matter effects but starts getting imperfect  in case of new physics scenarios. This is due to the fact that $\delta_{13}$ and new physics terms appear in a coupled way in the probability expressions.}, the  matter contributions are independent of $\delta_{13}$ (i.e., not arising due to the intrinsic CP phase, $\delta_{13}$, see Eq.~\ref{eq:mar} below) and above quantity is helpful to extract the intrinsic contribution~\cite{Marciano:2006uc}.

From Ref.~\cite{Marciano:2006uc}, one can analytically see that for $\nu_\mu \to \nu_e$ oscillation, 
\begin{equation}
\label{eq:mar}
\Delta P^{CP}_{\mu e} = \Delta P^{CP}_{\mu e} (\sin \delta_{13})  + \Delta P^{CP}_{\mu e} ({\textrm{matter}}, \delta_{13}{\textrm{-indep}}) + \ldots~,
\end{equation}
which implies decoupling of the intrinsic and extrinsic effects may be possible near the peak energy. 
Of course, the decoupling is not expected to work in general.

In Figs.~\ref{fig:9}-\ref{fig:11}, we show the oscillograms of \dacp~ in the plane of $\e$~and $\len$~ for the appearance and the disappearance channels. The three rows correspond to the different appearance or disappearance channels (as mentioned in the  subscripts of quantities plotted on the $y-$axes of the plots) while the columns are for SI, NSI and sterile cases. Unless otherwise stated, in this and the remaining plots, we take the additional phases in NSI and sterile cases to be zero for the sake of simplicity. These plots depict how new physics effects impact the inferences about intrinsic CP effects in the region in $\e-\len$~ plane.
Fig.~\ref{fig:11} is similar to Fig.~\ref{fig:9} but shows the T asymmetry.  

The following observations can be made in connection with the difference in CP asymmetries for the three appearance channels (see Fig.~\ref{fig:9}).

\begin{itemize}
\item $\nu_{\mu} \to \nu_{e}$ channel : 
\\
In the SI case, we note that the regions of large asymmetry ($\gtrsim 18\%$) are more concentrated at lower energies ($\lesssim 2$ GeV). Any experiment operating at higher energies and longer baselines cannot probe intrinsic CP violation via the quantity considered. The whitish region around a baseline of $\sim 8000$ km is due to the fact that it is near the magic baseline where the CP dependence vanishes~\cite{Barger:1980tf}. 
For NSI and sterile cases, we see the pattern in the oscillogram changes. However in case of NSI, the changes are more drastic. The new significant dark patches in the NSI case can be accounted for by assuming non-zero value of a particular NSI parameter (see Appendix~\ref{app_b}). Origin of different colours can be attributed to different parameters. For instance, in the $\nu_e\to\nu_ \mu$ and $\nu_e \to \nu_\tau$ channels, the orangish patch ($\sim 8-10\%$) that lies in $\e \in (5-9)$ GeV, $\len \in (300 - 2000)$ km is due to the presence of $\varepsilon_{ee}$. 
The sterile case is similar to SI with new features in the entire oscillogram plot. The wiggles arise due to the fast oscillations induced by  $\delta m^2_{41}$ which is large in comparison to the other mass squared splittings.
Among the considered experiments, T2K or T2HK seem to have the potential to extract the intrinsic CP phase from the  probability level discussion as far as the $\nu_{\mu} \to \nu_{e}$ channel is concrened.

\item $\nu_{\mu} \to \nu_{\tau}$ channel :
Using this channel, extracting intrinsic CP violation is hard as can be seen from large whitish regions in the oscillogram for the SI case. Again the pattern is very different for NSI and sterile cases. 
From the probability level oscillogram in the $\nu_\mu \to \nu_\tau$ channel, we note that in NSI case, DUNE lies on a darker patch. This gives an impression that in presence of NSI, extraction of intrinsic CP phase may be easier as compared to SI case. However this is misleading since the source of intrinsic CP violation remains the same in both cases (NSI phases are set to zero). 
The substructures cancel in sterile case due to the fact that the wiggles are independent of $\delta_{13}$ (see Fig.~\ref{fig:mt_analysis} in Appendix~\ref{app_c}).

\item $\nu_{e} \to \nu_{\tau}$ channel : 
The gross features are similar to $\nu_{\mu} \to \nu_{e}$ channel. The darker  regions can be understood from the size of the wiggles in Fig.~\ref{fig:mt_analysis} in Appendix~\ref{app_c}.

\end{itemize}

The following observations can be made in connection with CP plots for the three disappearance channels (see Fig.~\ref{fig:10}).

\begin{itemize}

\item $\nu_{e} \to \nu_{e}$ channel : \\
There is no $\delta_{13}$ dependence in the $\nu_e \to \nu_e$ channel~\cite{Akhmedov:2004ny} and hence the SI oscillogram is blank. NSI introduces 
significant effect in this channel. The features can be understood from Appendix~\ref{app_b}. Also, in sterile case, there are smaller dark patches as well as wiggles.

\item $\nu_{\mu} \to \nu_{\mu}$ channel :\\
Here the dependence on $\delta_{13}$ is mild for SI case~\cite{Akhmedov:2004ny}. Again, differences arise in case of NSI and sterile. 

\item $\nu_\tau \to \nu_\tau$ channel :\\
Here the dependence on $\delta_{13}$ is mild and similar to the case of $\nu_\mu \to \nu_\mu$ channel for SI case~\cite{Akhmedov:2004ny}. For NSI and sterile, we see darker regions and wiggles respectively. 

\end{itemize}

The following observations can be made in connection with T asymmetry plots for the three appearance channels (Fig.~\ref{fig:11}).  
The oscillograms show features similar in nature to the CP case  but there are fewer dark patches than CP case. Though the SI and 
NSI cases are indistinguishable, wiggles appear in sterile case.

\section{Implications for long baseline accelerator experiments}
\label{lbl}
%
Below we give a very brief description of the long baseline experiments considered to describe the implications of 
our results at the level of event rates. For more details, please see Table~\ref{tab:sys} (see also \cite{Masud:2016bvp}).
\begin{description}
 \item
{\bf {DUNE :}} We consider a design that uses {120 GeV} proton beam with a power of 
$1.0$ MW which corresponds to 
\begin{equation}
\dfrac{\textrm{P.O.T./year}}{[5.0 \times 10^{20}]} \sim \dfrac{{\textrm{Proton beam power }}}{[1 ~{\textrm{MW}}]} \times \dfrac{T}{[10^{7} ~\textrm{sec}]} \times \dfrac{[120 ~\textrm{GeV}]}{E_p} 
\end{equation}
  We assume 5 and 5 years of run time in neutrino and antineutrino modes respectively. The total exposure
 comes is around 350 kt.MW.years.

 \item
{\bf {NOvA (Numi Off-axis Neutrino Appearance) experiment :}} 
We consider a design that uses {120 GeV} proton beam with a power of 
$0.7$ MW which corresponds to 
\begin{equation}
\dfrac{\textrm{P.O.T./year}}{[3.0 \times 10^{20}]} \sim \dfrac{{\textrm{Proton beam power }}}{[0.7 ~{\textrm{MW}}]} \times \dfrac{T}{[10^{7} ~\textrm{sec}]} \times \dfrac{[120 ~\textrm{GeV}]}{E_p} 
\end{equation}
  We assume 
3  and 3 years of run time in neutrino and antineutrino modes respectively. The total exposure
 comes around 84 kt.MW.years.
  \item
{\bf{T2K (Tokai to Kamioka) experiment :}} 
We consider a design that uses {50 GeV} proton beam with a power of 
$0.770$ MW which corresponds to 
\begin{equation}
\dfrac{\textrm{P.O.T./year}}{[4.15 \times 10^{20}]} \sim \dfrac{{\textrm{Proton beam power }}}{[0.770 ~{\textrm{MW}}]} \times \dfrac{T}{[10^{7} ~\textrm{sec}]} \times \dfrac{[50~\textrm{GeV}]}{E_p} 
\end{equation}
  We assume 3 and 3 years of run time in neutrino and antineutrino modes respectively. The total exposure
 comes around  103.95 kt.MW.years.

\item
{\bf{T2HK (Tokai to Hyper Kamiokande) experiment :}}
We consider a design that uses {30 GeV} proton beam with a power of 
$7.5$ MW which corresponds to  
\begin{equation}
\dfrac{\textrm{P.O.T./year}}{[8.0 \times 10^{21}]} \sim \dfrac{{\textrm{Proton beam power }}}{[7.5~{\textrm{MW}}]} \times \dfrac{T}{[10^{7} ~\textrm{sec}]} \times \dfrac{[30~\textrm{GeV}]}{E_p} 
\end{equation}
  We assume 
3 and 1 year of run time in neutrino and antineutrino modes respectively. The total exposure
 comes around 16.8 Mt.MW.years.

\end{description}

\begin{table}[ht]
\centering
\begin{tabular}{ |l| l| }
\hline
{\bf{DUNE}} & {\bf{T2K}} \\
$E_{p} =2.5$ GeV, $L=1300$ km & $E_p = 0.6$ GeV, $L=295$ km\\
 Runtime (yr) = 5 $\nu$ + 5 $\bar \nu$ 
  & Runtime (yr) = 3 $\nu$ + 3 $\bar \nu$  \\
35 kton, LArTPC & 22.5 kton, WC \\
$\epsilon_{app}=80\%$, $\epsilon_{dis}=85\%$
& $\epsilon_{app}=50\%$, $\epsilon_{dis}=90\%$\\
{{$R_\mu=0.20/{\sqrt E}$, $R_e=0.15/{\sqrt E}$}} &{{$R_\mu=0.085/{\sqrt E}$, $R_e=0.085/{\sqrt E}$}} \\
$E\in[0.5-10.0]$ GeV, Bin width = 250 MeV & $E\in[0.4-1.2]$ GeV, Bin width = 40 MeV \\
\hline
{\bf{NOvA}} & {\bf{T2HK}}\\
$E_{p} =1.6$ GeV, $L=810$ km & $E_p = 0.6$ GeV, $L=295$ km\\
 Runtime (yr) = 3 $\nu$ + 3 $\bar \nu$ 
& Runtime (yr) = 1 $\nu$ + 3 $\bar \nu$
\\
14 kton, TASD  & 560 kton, WC \\
$\epsilon_{app}=55\%$, $\epsilon_{dis}=85\%$
& $\epsilon_{app}=50\%$, $\epsilon_{dis}=90\%$
\\
{{$R_\mu=0.06/{\sqrt E}$, $R_e=0.085/{\sqrt E}$ }} 
& 
{{$R_\mu=0.085/{\sqrt E}$, $R_e=0.085/{\sqrt E}$}}
\\
$E\in[0.5-4.0]$ GeV, Bin width = 125 MeV & $E\in[0.4-1.2]$ GeV, Bin width = 40 MeV\\
\hline
\end{tabular}
\caption{\label{tab:sys} 
Detector configuration, efficiencies, resolutions and relevant energy ranges for \dune, \nova, \ttok, \ttohk.  }
\end{table}

\begin{figure}[htb]
\centering
\includegraphics[width=1.0\textwidth] 
{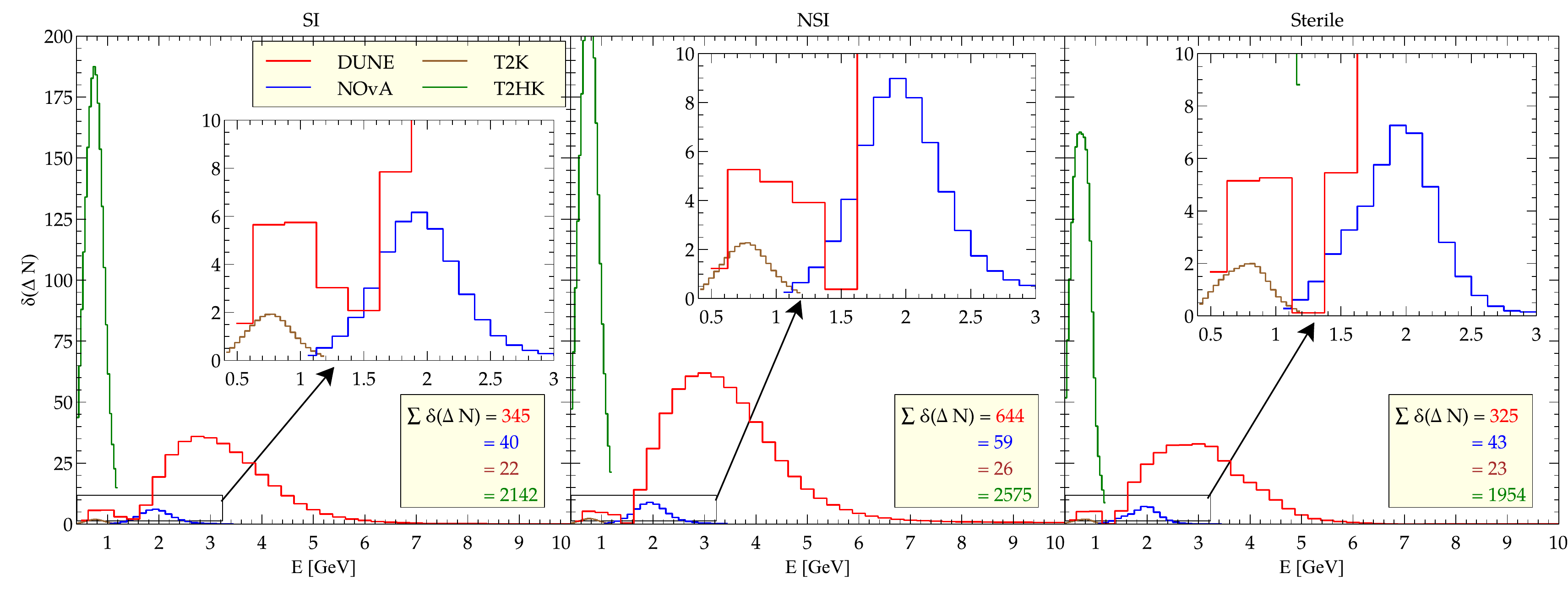}
\caption{\footnotesize{
$|\delta [\Delta N^{CP}_{\mu e}]|$ plotted as a function of $E$. The binning for the four experiments is different. 
 }}
\label{fig:14a}
\end{figure}

\begin{figure}[htb]
\centering
\includegraphics[width=1.0\textwidth] 
{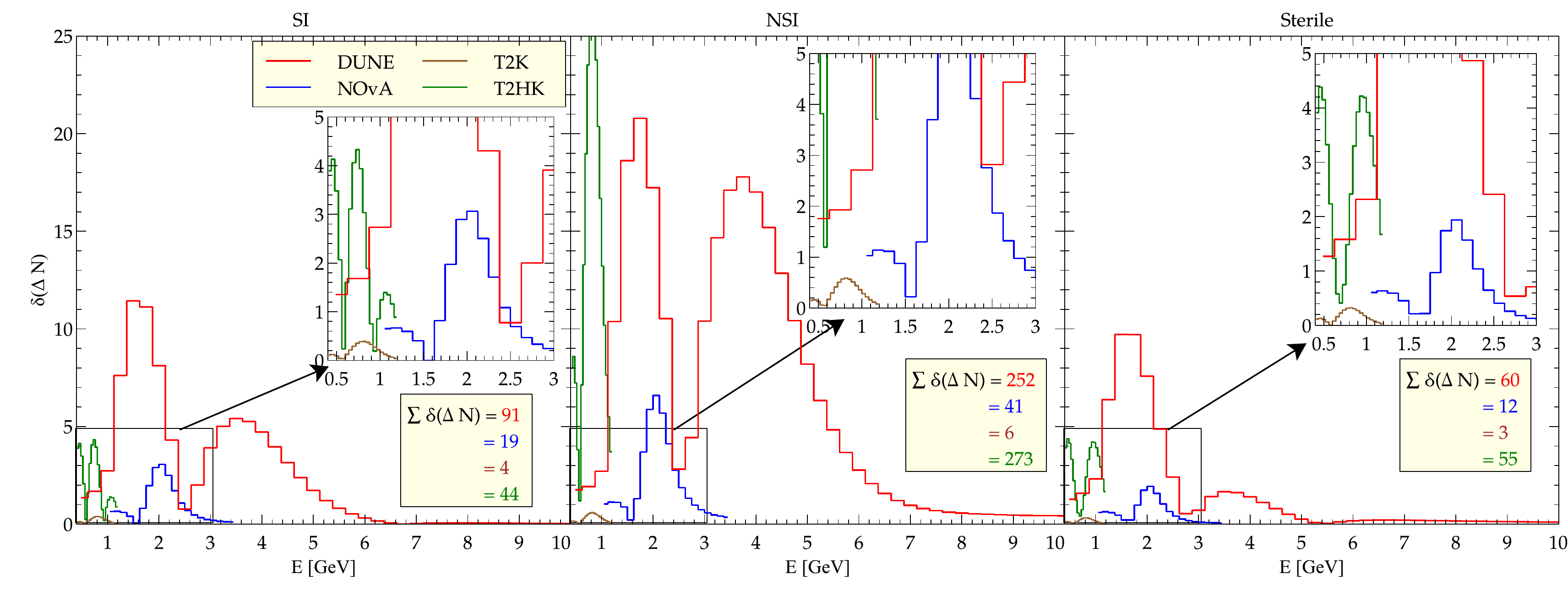}
\caption{\footnotesize{
$|\delta [\Delta N^{CP}_{\mu\mu}]|$ plotted as a function of $E$. The binning for the four experiments is different. 
 }}
\label{fig:14b}
\end{figure}

\begin{figure}[htb]
\centering
\includegraphics[width=1.0\textwidth] 
{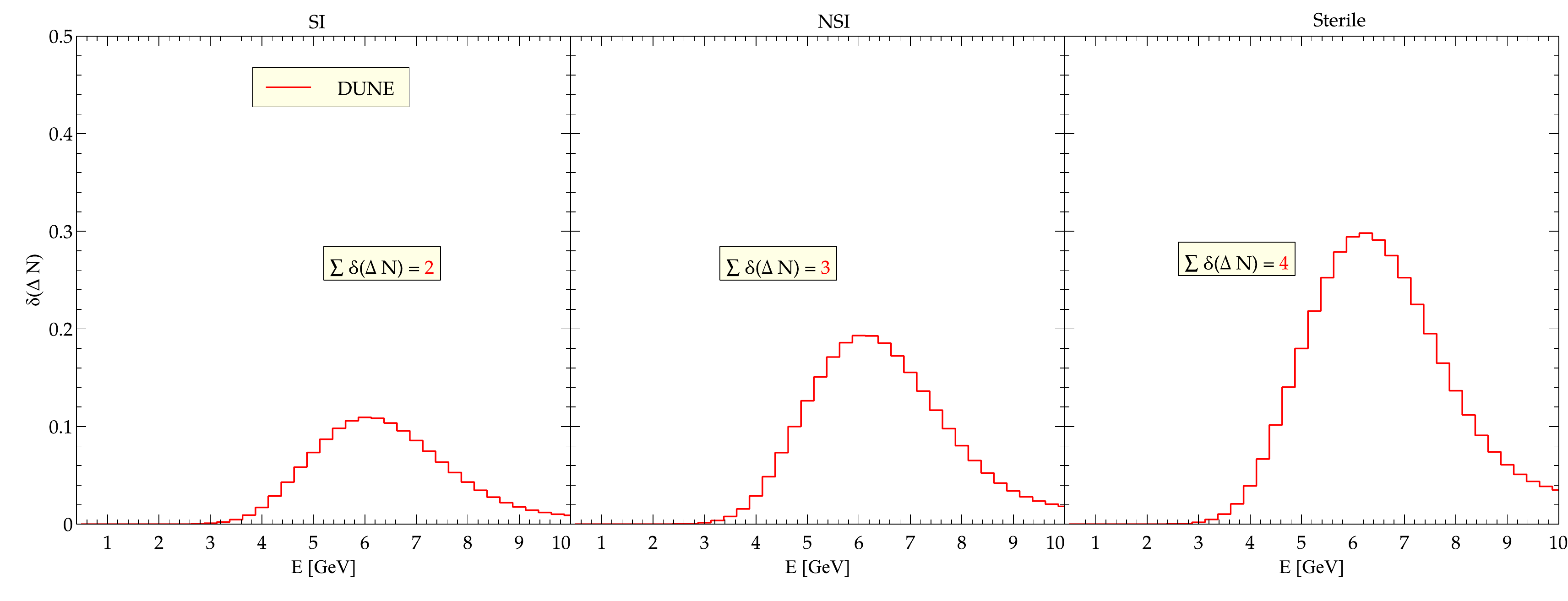}
\caption{\footnotesize{
$|\delta [\Delta N^{CP}_{\mu \tau}]|$ plotted as a function of $E$. We get handful events for DUNE and none for the  rest.  
 }}
\label{fig:14c}
\end{figure}

\begin{table}[htb]
\centering
\begin{tabular}{|l | c c | c c | c c |}
\hline
Experiment & \multicolumn{2}{c |}{SI} & \multicolumn{2}{c 
|}{NSI} & \multicolumn{2}{c 
|}{Sterile}  \\
                                         \cline{2-7}
                                        & $\nu_\mu \to \nu_e$ 
                                        & 
                                         $\nu_\mu \to \nu_\mu$ & 
                                         $\nu_\mu \to \nu_e$ &
                                          $\nu_\mu \to \nu_\mu$ &
                                          $\nu_\mu \to \nu_e$ &
                                          $\nu_\mu \to \nu_\mu$                                                                                   
                                        \\ 
\hline
&&&&&&\\
{{\dune} (NH, $\pi/2$)} & 345 & 91 & 644 & 252 & 325 & 60\\ 
{{\dune} (NH, $-\pi/2$)} & 328 & 91 & 261 & 252 & 197 & 60\\
{{\dune} (IH, $\pi/2$)} & 231 & 41 & 141 & 187 & 227 & 28\\ 
{{\dune} (IH, $-\pi/2$)} & 310 & 41 & 514 & 187 & 278 & 28\\ \hline
{{\nova}} (NH, $\pi/2$) & 40 & 19 & 59 & 41 & 43 & 12\\ 
{{\nova}} (NH, $-\pi/2$) & 34 & 19 & 16 & 41 & 22 & 12\\
{{\nova}} (IH, $\pi/2$) & 33 & 12 & 15 & 36 & 38 & 9\\ 
{{\nova}} (IH, $-\pi/2$) & 36 & 12 & 53 & 36 & 43 & 9\\ \hline
{{T2K}} (NH, $\pi/2$) & 22 & 4 & 26 & 6 & 23 & 3\\ 
{{T2K}} (NH, $-\pi/2$) & 15 & 4 & 11 & 6 & 8 & 3\\
{{T2K}} (IH, $\pi/2$) & 14 & 3 & 11 & 5 & 22 & 3\\ 
{{T2K}} (IH, $-\pi/2$) & 21 & 3 & 24 & 5 & 21 & 3\\ \hline
{{T2HK}} (NH, $\pi/2$) & 2142 & 44 & 2575 & 273 & 1954 & 55\\ 
{{T2HK}} (NH, $-\pi/2$) & 2001 & 44 & 1567 & 273 & 1492 & 55\\
{{T2HK}} (IH, $\pi/2$) & 2021 & 32 & 1574 & 217 & 2552 & 71\\ 
{{T2HK}} (IH, $-\pi/2$) & 2094 & 32 & 2531 & 217 & 2658 & 71 \\ \hline
\end{tabular}
\caption{\label{tab:events} 
$|\delta [\Delta N^{CP}_{\alpha\beta}]|  = |[\Delta N_{\alpha\beta}^{CP}] (\delta_{13} = \pm \pi/2) - [\Delta N_{
\alpha\beta}^{CP}] (\delta_{13}=0)|$ summed over energy bins for different experiments for NH and IH.
For NSI, we show the collective case when the NSI parameters  $|\eem|=0.04, |\eet|=0.04, |\emt|=0.03, 
\emm=0.06, \ett=0.1,  \eee=0.4$, $\varphi_{e\mu}=0$, $\varphi_{e\tau}=0$, $\varphi_{\mu\tau}=0$ are 
considered. The sterile parameters are as mentioned in Table~\ref{tab:parameters_sterile}. }
\end{table}

We present our results at the level of event rates using the following quantity 
\bea
\delta [\Delta N^{CP}_{\alpha\beta}]  = [\Delta N_{\alpha\beta}^{CP}] (\delta_{13} = \pi/2) - [\Delta N_{
\alpha\beta}^{CP}] (\delta_{13}=0)~.
\label{eq:Easymm}
\eea
Here the first term on the RHS corresponds to the case of maximal CP violation ($\delta_{13} = \pi/2$) 
while the second term corresponds to CP conservation ($\delta _{13} = 0$). Since all the accelerator 
experiments mentioned above can produce $\nu_\mu$ only at the source (pion decay), we discuss the 
implications of our results at the level of event rates using $\nu_\mu \to 
{\textrm{any flavour}}$. Note that the binning and energy range of the experiments are different (see Table.~\ref{tab:sys}). 

\begin{itemize}

\item $\nu_\mu \to \nu_e$ : Among all the considered experiments,  the total event rate is highest for T2HK by a 
large margin. This high statistics is due to large detector size 
and beam power. Moreover, this means that one would be able to disentangle intrinsic and extrinsic sources of 
CP violation better with T2HK. The much shorter baseline ensures that matter effects are  small which in turn 
simplifies the extraction of intrinsic CP phase. 
We can note that NSI and sterile scenarios also retain this feature as long as additional phases are set to zero 
(see Table~\ref{tab:events} and Fig.~\ref{fig:14a}).

\item $\nu_\mu \to \nu_\mu$  : Here, in case of SI, DUNE seems to be the best choice. But, in presence of new 
physics such as NSI or sterile, T2HK seems to do slightly better though DUNE is also competitive 
(see Table~\ref{tab:events} and Fig.~\ref{fig:14b}). 

\item $\nu_\mu \to \nu_\tau$ : The number of events are scarce even after using a higher energy beam tune 
and evidently we can not draw useful conclusions from this channel (see Table~\ref{tab:events} and Fig.~\ref{fig:14c}). 
\end{itemize}

\section{Concluding remarks}
\label{con}

It is fair to say that neutrino oscillation physics has entered a precision era and the upcoming long baseline experiments are expected to shed some light on one of 
 the crucial unknown parameters in the oscillation framework - the leptonic CP phase. Going beyond the recent studies revealing how potential new 
 physics scenarios can hinder the clean determination of this important parameter~\cite{Masud:2015xva,
deGouvea:2015ndi,Coloma:2015kiu,Masud:2016bvp,Masud:2016gcl,Blennow:2016etl,Forero:2016cmb,
deGouvea:2016pom,Fukasawa:2016lew,Liao:2016orc,Gandhi:2015xza,Agarwalla:2016mrc,Agarwalla:2016xxa,Choubey:2016fpi,Dutta:2016glq,Agarwalla:2016xlg,Blennow:2016jkn,Deepthi:2016erc}, we address the issue of clean separation of the 
 intrinsic leptonic CP phase from the extrinsic contribution arising due to SI as well as new physics\footnote{{{We assume that only source of intrinsic CP violation is due to $\delta_{13}$ which is very optimistic. In principle, the
  new physics scenarios considered here can also bring in more sources of intrinsic CP violation via pure phase terms. }}}. We also  show the impact of new physics on testing non-unitarity.  
  
  Accelerator-based long baseline experiments are plagued with the problem of clean  separation of intrinsic and extrinsic CP violating terms due to the fact that neutrinos propagate through Earth matter.  Resolution of this is a difficult task and there are several suggestions including new observables in literature.   In order to elucidate and quantify our results, we  use an  observable quantity given by Eq.~\ref{eq:asymmdif} (see~\cite{Arafune:1997hd,Nunokawa:2007qh,Ohlsson:2013ip}) and scan the range of energies and path lengths relevant to long baseline experiments.  We consider two new physics scenarios - NSI in propagation and additional sterile neutrinos.   We depict our outcome in the form of oscillogram plots in $\len-\e$ space where we unravel the regions where the impact of new physics on the oscillograms could be potentially
  large. We also  highlight the differences in the different physics cases for some benchmark values of new physics parameters.  Our discussion is mostly targeted towards accelerator based neutrino experiments with $\lbye \sim 500$ \km/\gev~ but can easily be extended to short baseline experiments and very long baseline experiments.  Other than the question of separating the intrinsic CP contribution, we also discuss the impact of additional sterile neutrinos on the unitarity conditions.
   
For the sake of simplicity, we set the additional intrinsic CP phases (induced by new physics) to zero and discuss the impact of additional parameters appearing in NSI and sterile cases that are extrinsic in nature.  Furthur through the event rates (see Eq.~\ref{eq:Easymm}) for realistic configurations for some of the ongoing and planned long baseline experiments, we show which experiment has better potential to answer the questions that we have posed in this article. 
We have considered four long baseline experiments - T2K, T2HK, DUNE and NoVA and also taken into account both the appearance ($\nu_\mu \to \nu_e$) and disappearance ($\nu_\mu \to \nu_\mu$) channels for each of them (see Sec.~\ref{lbl}). We demonstrate how even the restricted class of CP conserving new physics effects complicate the separation of intrinsic CP phase from the extrinsic CP effects that can come from SI or new physics. Our main results can be summarised as follows :

\begin{itemize}

\item {\sl Non-unitarity :}
Deviation from unitarity (in the sterile case) at the probability level can be discernible by looking at various  $\e$~and $\len$~ ranges where one gets darker regions.  
{{Fig.~\ref{fig:9m}  shows the deviation from unitarity for three different values of $\delta_{13}$. Obviously, in vacuum, one would expect to get blank region whenever the source of CP violating phase  vanishes ($\delta_{13}$ being the only source of CP violation, see Eq.~\ref{unitary}). This corresponds to the middle column of  Fig.~\ref{fig:9m}. Instead we see some pattern even for $\delta_{13}=0$ and this can be attributed to 
the SI with matter which contributes to non-unitarity.  }}
   We can note that only in channels involving  $\nu_\tau$, it might be possible for DUNE or NOvA or T2HK to reveal some signature of non-unitarity.  
However this is not expected to be useful at the level of events.  Non-unitarity is very hard to probe in the $\nu_\tau$ channel using any of the long baseline experiments primarily because one is statistically limited in case of tau events (see Fig.~\ref{fig:9m}).

\item {\sl Extraction of intrinsic CP violating component and comparison of new physics scenarios with SI :} 
At the probability level, the darker shaded regions imply larger  influence of new physics. This dark region should not be thought of as aiding the extraction of intrinsic CP component in any given channel, rather it makes the situation more complicated. Some of the ongoing and future experiments are shown as bulleted points along the curve representing the first oscillation maximum. For lower values of $\e$~and $\len$, it is expected that the NSI effects would be small and hence one could in principle have a clean detection of intrinsic component. From the oscillograms, we can note that the impact of new physics scenarios is more prominent at larger values of $\e$~and $\len$. 
Also, note that NoVA and DUNE lie on lighter shaded region of the oscillogram while T2K (and T2HK) is at a darker patch. 
Hence the  baseline choice of T2K or T2HK is desirable  in order to extract the intrinsic component from the probability level analysis.  Finally, at the level 
of events, T2HK wins due to the large statistics  in order to cleanly extract the intrinsic contribution (see Fig.~\ref{fig:9} - \ref{fig:11}).

\item {\sl Event analysis :} At the level of event rates, we find that T2HK offers best statistics among all  the considered experiments in case of $\nu_{\mu} \to \nu_{ e}$ channel. But, DUNE is competitive with T2HK if we consider $\nu_{\mu} \to \nu_{\mu}$ channel.  Tau appearance channel is mostly not useful due to limited statistics (see Figs.~\ref{fig:14a} - \ref{fig:14c} and Table~\ref{tab:events}).

\end{itemize}

Finally some comments concerning the validity of our approach are in order.  We assume that only source of intrinsic CP violation is due to $\delta_{13}$ which is very optimistic. In principle, the  new physics scenarios considered here can also bring in more sources of intrinsic CP violation via pure phase terms.
Any source of new physics therefore has both intrinsic (i.e. phases) and extrinsic components and discussing the problem with both components is rather cumbersome. 
In fact, the separation of intrinsic contribution using a quantity like $\delta (\Delta P^{CP}_{\alpha\beta})$  is feasible only when there is one source of intrinsic CP violation ($\delta_{13}$) present. For a more general scenario with phases introduced in the new physics sector, one needs to think of appropriate observables to be able to separate out the intrinsic contribution. 

Nonetheless we would like to stress that our overall approach to survey the impact of CP conserving new physics scenarios is quite general and can be applied to other new physics scenarios or other regimes in $\e -\len$~space. The discussion in the present work is targeted towards accelerator-based neutrino experiments with $\lbye \sim 500$ $\km/\gev$ but the ideas can easily be extended to short baseline experiments or very long baseline experiments.

\section*{Acknowledgements} 
It is a pleasure to thank Raj Gandhi for useful discussions and critical comments on 
the manuscript.  
 We acknowledge the use of HRI cluster facility to carry out computations in this work. 
 JR acknowledges financial support in the form of a research fellowship from UGC-BSR 
 (Ref. No.F.25-1/2013-14(BSR)/7-95/2007(BSR)).
   MM would like to thank JNU and Utpal Chattopadhyay at Indian Association for 
   Cultivation of Science, Kolkata for academic visits and support from the DAE neutrino 
   project at HRI during the progress of the present work. The work of MM was funded
    by the Spanish grants FPA2014-58183-P, SEV-2014-0398 (MINECO) and 
    PROMETEOII/2014/084 (Generalitat Valenciana).
    MM and PM would like to thank HRI for a visit during the finishing stages of this work. 
    PM acknowledges support from University Grants Commission under the second phase of University with Potential of Excellence at JNU and DST-PURSE 
    grant at JNU as well as partial support from the European Unions Horizon 2020 research and innovation programme under Marie Sklodowska-Curie grant No 674896. 
    We would like to thank the anonymous referees for constructive suggestions. 
    %


\begin{appendix}
\numberwithin{equation}{subsection}
\appendix
\section*{Appendices}
\renewcommand{\thesubsection}{\Alph{subsection}}
%
\subsection{Origin of oscillogram pattern depicting non-unitarity in the Sterile case}
\label{app_a}

\begin{figure}[htb]
\centering
\includegraphics[width=\textwidth]
{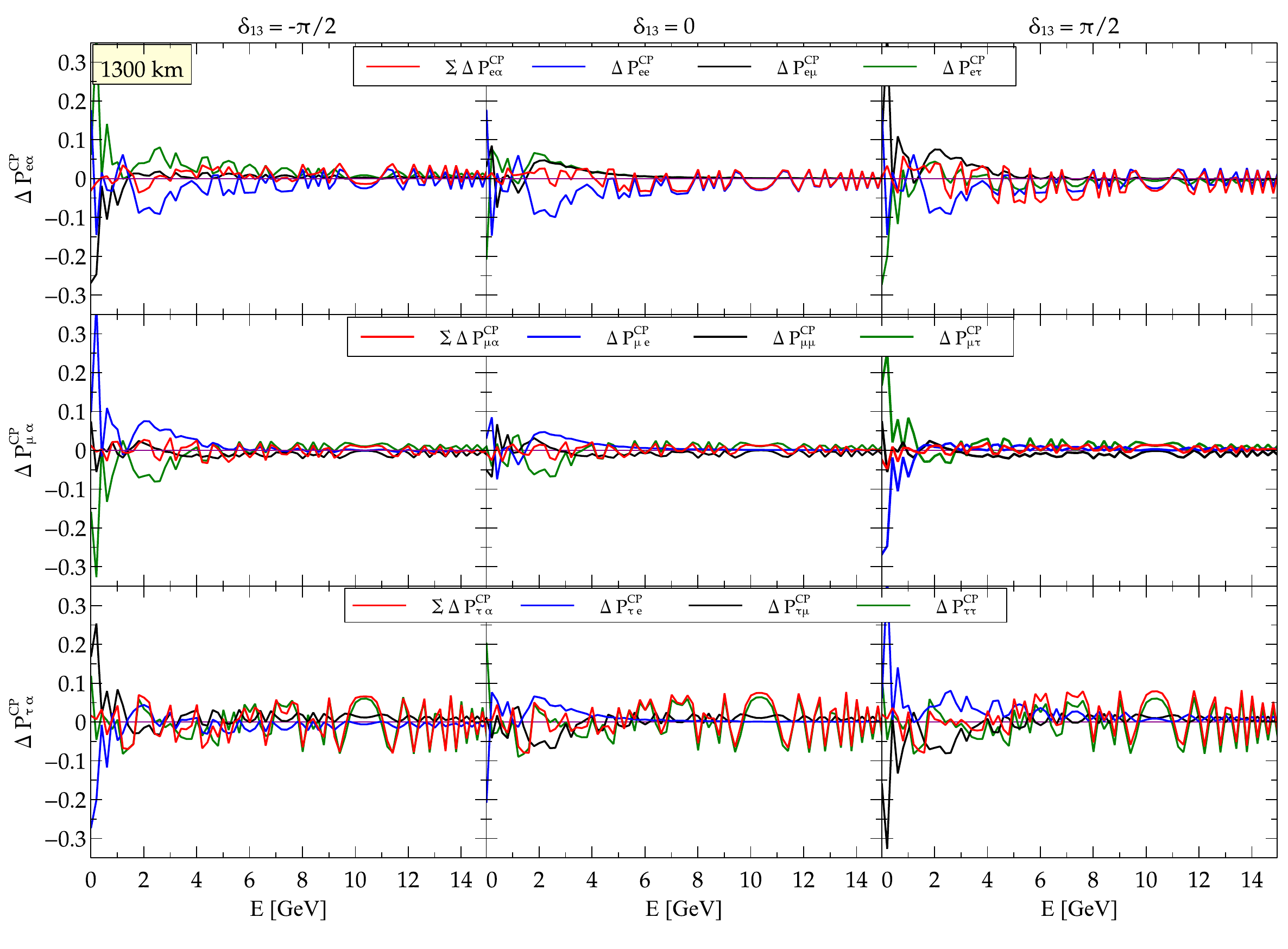}
\caption{\footnotesize{$\Delta P^{CP}_{e\alpha}$, $\Delta P^{CP}_{\mu\alpha}$ and $\Delta P^{CP}_{\tau\alpha}$ plotted as a function of $E$ for a fixed baseline of 1300 km. 
 }}
\label{fig:nu_analysis}
\end{figure}

In order to explain the features of different panels in Fig.~\ref{fig:9m}, in Fig.~\ref{fig:nu_analysis} we show the individual components (\textcolor{blue}{blue}, \textcolor{black}{black} and \textcolor{darkgreen}{darkgreen} curves) and the sum of the contributions in each row (\textcolor{red}{red} curves) appearing in Fig.~\ref{fig:9m} for a fixed baseline of 1300 km.  As we can see the \textcolor{red}{red} curve is rapidly oscillating which leads to thin light and dark patches in Fig.~\ref{fig:9m} along the horizontal line at 1300 km. The amplitude of the \textcolor{red}{red} curve depends on the value of the sterile mixing angle relevant in each channel (see the main text).  

\subsection{Origin of dark regions in the CP and T oscillograms in NSI case}
\label{app_b}

The approximate analytic expressions for probabilities upto second order in small parameters ($\hat{r_A}, s_{13}, \varepsilon$'s) in different channels in case of NSI are given in~\cite{Kikuchi:2008vq,Asano:2011nj,Liao:2016hsa}. Using the analytic expressions, we attempt to explain the distinct features of the oscillograms (Figs.\ \ref{fig:9}, \ref{fig:10} and \ref{fig:11}). In order to simplify the tedious expressions, we assume the following
\begin{itemize}
 \item Normal hierarchy (NH)
 \item  $\eema = \eeta$ which is consistent with our choice of parameters in generating the oscillograms. This results in the cancellation of terms $\propto (\eema - \eeta)$
 and allows for useful simplifications in the analytical formulae.
 
 \item the NSI phases  are set to zero ($\eemp = \eetp = \emtp = 0$)
 
 \item  Upto second order, the expression for $P(\nu_{e} \rightarrow \nu_{\alpha})$ (where, $\alpha = e,\mu,\tau$) contain the NSI parameters $\eem, \eet, \eee$~
 \cite{Kikuchi:2008vq}. Hence, the mild effect of $ \emt, \emm, \ett$ on electron sector cannot be understood from these. 
  \item To get the anti-neutrino probabilities, one needs to do the following replacements 
  $r_A \to -r_A$, $\delta_{13} \to -\delta_{13}, 
  \varepsilon_{\alpha\beta} \to \varepsilon_{\alpha\beta}^{*}$.

 \end{itemize}
 
  In order to facilitate the presentation, we define the following quantities (the \textit{bars} above indicate the corresponding quantities for 
  antineutrinos.).
  \begin{enumerate}
  \item
    \bea \label{eq:a_del}
  r_A &=& \frac{A}{\ldm} \approx 0.03 ~E [GeV]~\rho[gm/cc]~, \nonumber\\
  \hat{r_A} &=& r_A \left(1+\eee \right), \nonumber \\
  \lambda  &=& \dfrac{\ldm }{2 E}~.
  \eea
  
  \item 
  \bea \label{eq:cd}
  C &=& \frac{\hat{r_A}}{\sqrt{2}}(\eema + \eeta)~; \bar{C} = -C \nonumber\\
  D_{1} &=& \frac{\sin((1-\hat{r_A})\lambda L/2)}{1-\hat{r_A}} - \frac{\sin((1+\hat{r_A})\lambda L/2)}{1+\hat{r_A}}~; \bar{D_{1}} = -D_{1} \nonumber\\
  D_{2} &=& \frac{\sin((1-\hat{r_A})\lambda L/2)}{1-\hat{r_A}} + \frac{\sin((1+\hat{r_A})\lambda L/2)}{1+\hat{r_A}}~; \bar{D_2} = D_2 \nonumber\\
  D &=& \frac{\sin((1-\hat{r_A})\lambda L/2)}{(1-\hat{r_A})^{2}} - \frac{\sin((1+\hat{r_A})\Delta)}{(1+\hat{r_A})^{2}}~; \bar{D} = -D
  \eea
  
 \item
 \bea\label{eq:omega}
 \Omega &=& |\Omega| e^{i\omega}~ \text{\quad where \quad} \nonumber \\
 |\Omega| &\approx& \sqrt{\frac{s_{13}^{2} + C^{2} + 2s_{13}C\cos\delta_{13}}{\hat{r_A}^{2}}}~, \nonumber \\
 \tan\omega &=& \frac{C\sin\delta_{13}}{s_{13} + C\cos\delta_{13}}~, \nonumber \\
 \text{and }
 |\bar{\Omega}| &\approx& \sqrt{\frac{s_{13}^{2} + C^{2} - 2s_{13}C\cos\delta_{13}}{\hat{r_A}^{2}}}~, \nonumber \\
  \tan{\bar{\omega}} &=& \frac{C\sin\delta_{13}}{s_{13} - C\cos\delta_{13}}~.
 \eea
 Note that, $\omega$ vanishes at $\delta_{13} =0$.

  \item
  \bea\label{eq:sigma}
  \Sigma &=& |\Sigma|~\exp\{i\sigma\}~\textrm{\quad where \quad} \nonumber \\
  |\Sigma| &=& (r_\lambda/2r_A) \sin2\theta_{12} + (\propto (\eema - \eeta)) 
  \approx (r_\lambda/2r_A) \sin2\theta_{12}~,\nonumber \\
  \sigma &\simeq& \delta_{13} \text{\quad for $\varphi_{\alpha \beta} = 0$~.}
  \eea
  Hence, $\bar{|\Sigma|} = -|\Sigma|$ and $\bar{\sigma} = -\sigma$.
  
  \end{enumerate}
  
Now we give the simplified expressions for the different sectors below. 

\subsection*{$\mu-\tau$ sector:}

\bea\label{eq:delta_mt}
\delta \Delta P^{CP}_{\mu\tau} &=& \delta(P_{\mu\tau} - \bar{P}_{\mu\tau}) \nonumber\\
 &=& (P_{\mu\tau} - \bar{P}_{\mu\tau})|_{\delta_{13} = \pi/2} - (P_{\mu\tau} - \bar{P}_{\mu\tau})|_{\delta_{13} = 0} \nonumber\\
&\approx& 4s_{13}C\sin \lambda L/2 \left\{\cos(\hat{r_A} \lambda L/2)D-\lambda L/2(\cos \lambda L/2)\frac{2r_A}{1-\hat{r_A}^{2}}\right\} \nonumber\\
&+& r_\lambda \sin2\theta_{12} \sin(\hat{r_A}\lambda L/2) \sin \lambda L/2
\sin\theta_{13}D_{1}~.
\eea
\begin{figure}[htb]
\centering
\includegraphics[width=\textwidth]
{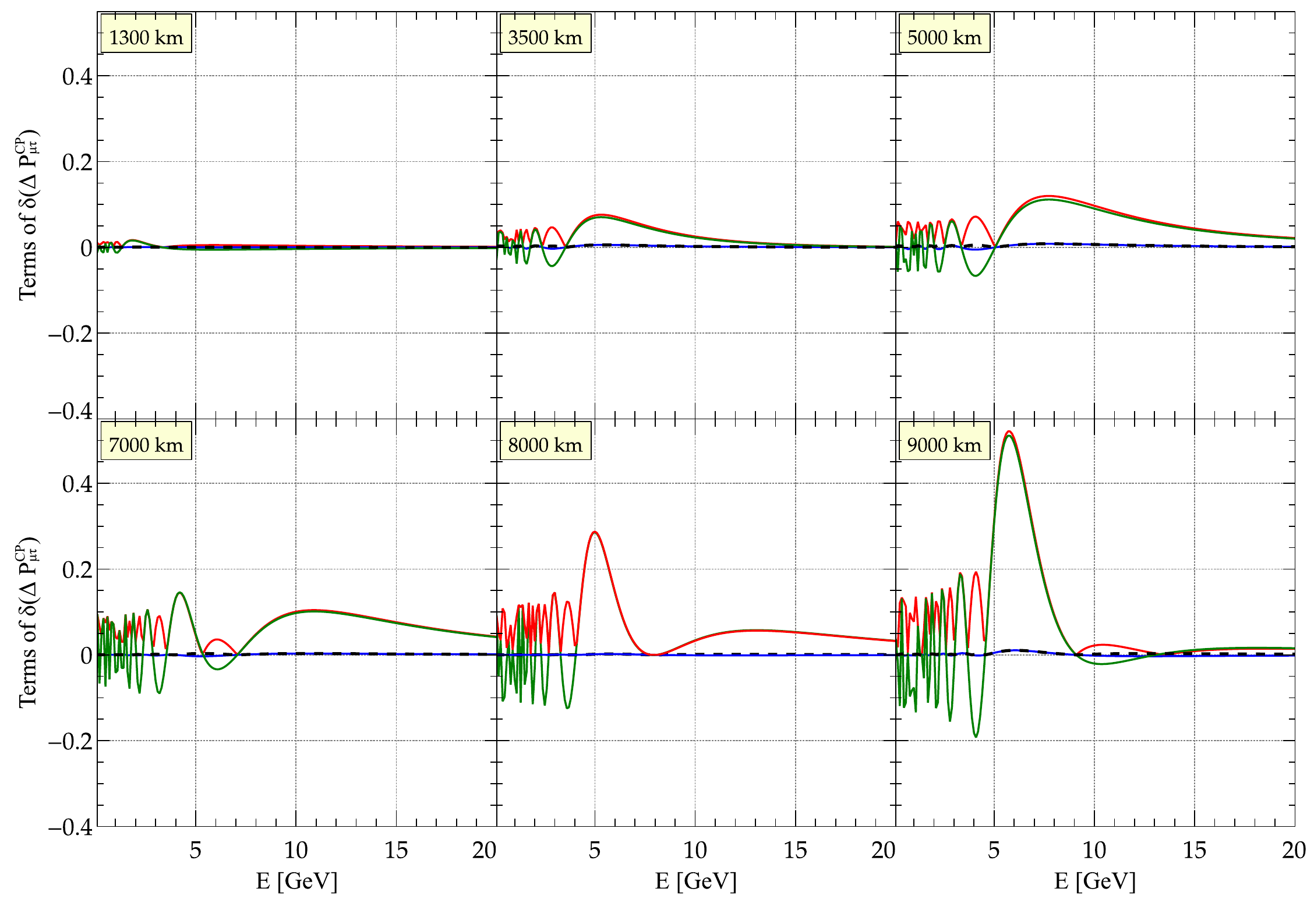}
\caption{\footnotesize{$\delta(\Delta P^{CP}_{\mu\tau})$ as a function of E[GeV] for 
6 fixed values of the baseline L[km]. The \textcolor{darkgreen}{darkgreen} (\textcolor{blue}{blue}) curve corresponds to the \textcolor{darkgreen}{first} (\textcolor{blue}{second} term of Eq.~(\ref{eq:delta_mt})). The \textcolor{red}{red} curve is the value of $|\delta \Delta P^{CP}_{\mu\tau}|$ in Eq.~(\ref{eq:delta_mt}). The black dashed curve corresponds to the value of $|\delta \Delta P^{CP}_{\mu\tau}|$ in the SI case.
 }}
\label{fig:mt_analysis}
\end{figure}

\bea\label{eq:delta_mm}
\delta \Delta P^{CP}_{\mu\mu} &=& \delta(P_{\mu\mu} - \bar{P}_{\mu\mu}) \nonumber\\
&=& (P_{\mu\mu} - \bar{P}_{\mu\mu})|_{\delta_{13} = \pi/2} - (P_{\mu\mu} - \bar{P}_{\mu\mu})|_{\delta_{13} = 0} \nonumber\\
&\approx& 4s_{13}C\bigg(D_{1}D_{2} + \frac{\hat{r_A}\lambda L/2 \sin(\lambda L)}{1-\hat{r_A}^{2}}
- D\cos(\hat{r_A}\lambda L/2)\sin\lambda L/2 \bigg) \nonumber\\
&+& 2 r_\lambda \sin2\theta_{12} \cos\lambda L/2 \frac{\sin(\hat{r_A}\lambda L/2)}{\hat{r_A}} 
\bigg[s_{13}D_{2} + \frac{2C\sin((1+\hat{r_A})\lambda L/2)}{1+\hat{r_A}}\bigg]
\eea

\begin{figure}[htb]
\centering
\includegraphics[width=\textwidth]
{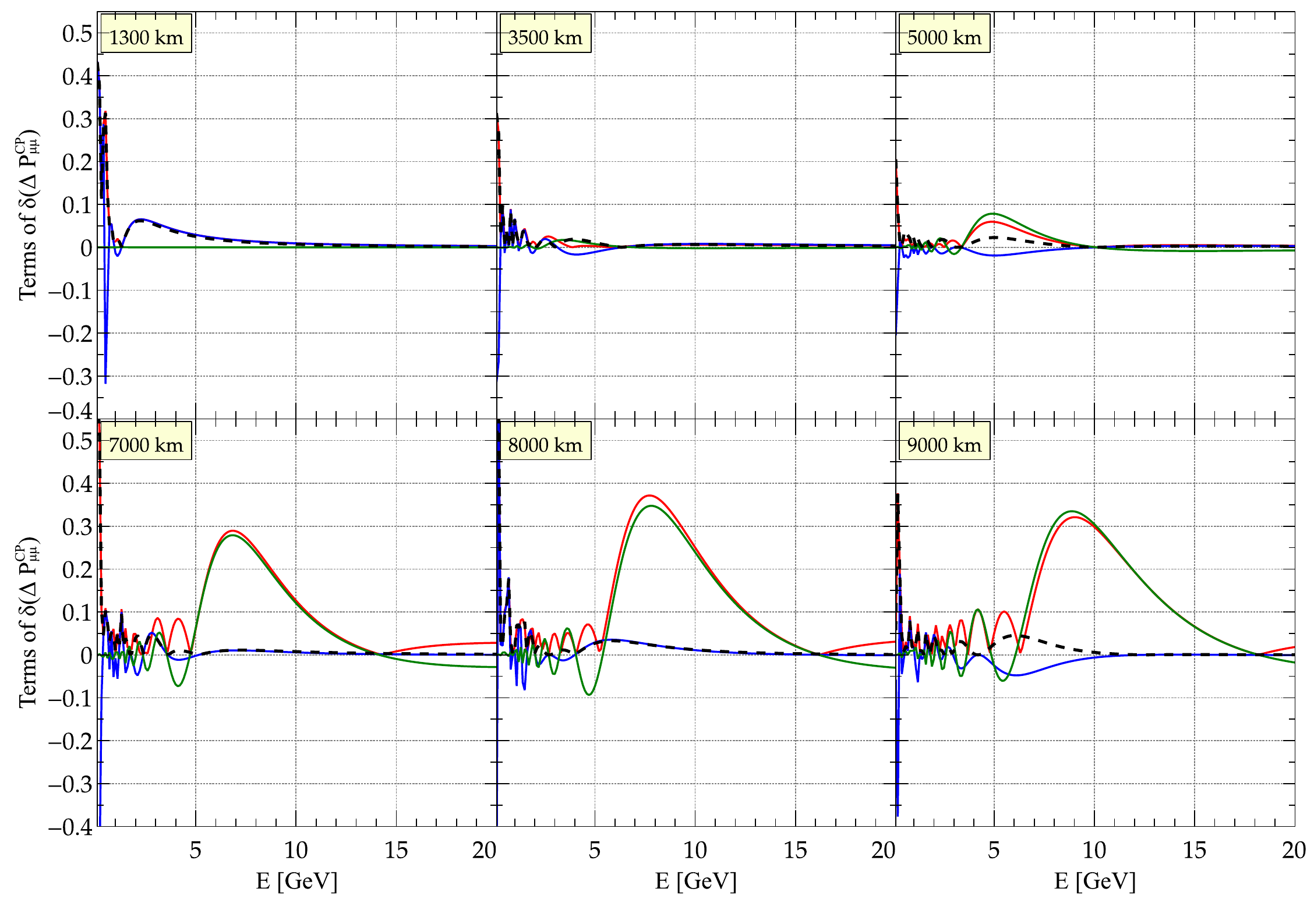}
\caption{\footnotesize{$\delta(\Delta P^{CP}_{\mu\mu})$ as a function of E [GeV] for 
6 fixed values of the baseline L [km]. The 
\textcolor{darkgreen}{darkgreen} (\textcolor{blue}{blue}) curve corresponds to the \textcolor{darkgreen}{first} (\textcolor{blue}{second} term of Eq.~(\ref{eq:delta_mm})). The \textcolor{red}{red} curve is the value of $|\delta \Delta P^{CP}_{\mu\mu}|$ in Eq.~(\ref{eq:delta_mm}). The black dashed curve corresponds to the value of $|\delta \Delta P^{CP}_{\mu\mu}|$ in the SI case.
 }}
\label{fig:mm_analysis}
\end{figure}

These expressions serve to explain the qualitative features obtained in Fig.~\ref{fig:9} and \ref{fig:10}. We note that $\delta \Delta P^{CP}_{\mu\tau}$ 
 and $\delta \Delta P^{CP}_{\mu\mu}$  are shown in the middle row of Fig.~6 and 7 respectively.  
In Figs.~\ref{fig:mt_analysis} and \ref{fig:mm_analysis}, different terms in Eq.~(\ref{eq:delta_mt}) and 
(\ref{eq:delta_mm}) have been plotted respectively and we can connect these plots with Fig.~\ref{fig:9} and \ref{fig:10}.  
 We observe the following distinct features from Figs.~\ref{fig:mt_analysis} 

\begin{itemize}
 \item The gross nature of $|\delta \Delta P^{CP}_{\mu\tau}|$ and $|\delta \Delta P^{CP}_{\mu\mu}|$ (the \textcolor{red}{red} curves) is mostly dictated by the \textcolor{darkgreen}{first term}  ($\propto C s_{13}$)
  in Eq.~(\ref{eq:delta_mt}) and (\ref{eq:delta_mm}) respectively.  
 The first term is purely NSI term and is the dominant term in the expression. 
Note that the  \textcolor{blue}{second term} in Eq.~(\ref{eq:delta_mt}) and (\ref{eq:delta_mm}) is scaled by  $r_\lambda$ 
 $(\approx 10^{-2})$ which is small in comparison to the first term.

 \item Let us compare the plots at different baselines.  For shorter baselines, $|\delta \Delta P^{CP}_{\mu\tau}|$ is insignificant for all values of energies but  for some choice of energies it becomes prominent as the baseline increases. This prominence can be visualized as a series of peaks in the plot.  As the baseline increases, these peaks show the following tendencies - shift towards right, becoming broad or narrow and change in prominence (amplitude) among the different peaks. 
There are two prominent long dark orangish stretches in Fig.~\ref{fig:9} - one  around $E \sim 3.5-7.5$ GeV and $L \sim 1000 - 5000 km$  and another thinner one around $E \sim 2.5-3.5$ GeV and $L \sim 1000 - 5000~ km$. These can be explained from the first (right most) peak in Fig.~\ref{fig:mt_analysis}. The slant of these stretches is due to shift in the  peak position towards right as the baseline increases. The \textit{sharpness} of the second peak of Fig.~\ref{fig:mt_analysis} and its relatively mild shift from $\sim 3 ~GeV$ to $\sim 6 ~ GeV$ as $L$ increases from $\sim 3500 ~km$ to $\sim 9000 ~km$ produces the less slanted thin dark stretch in Fig.\ \ref{fig:9}.
   In addition, there are two  dark patches at very long baselines in Fig.~\ref{fig:9}  around $E \sim 3-5$ GeV and $L \sim 8000-10000 ~km$. The sudden rise in magnitude of the second peak at around $\gsim 8000~km$  produces the two dark spots in Fig.~\ref{fig:9} at longer baseline values.

\item The features in Fig.~\ref{fig:mm_analysis} are grossly similar to Fig.~\ref{fig:mt_analysis}. 

The peaks can be mapped to the dark patches/regions in the NSI plot (middle row and middle panel) of Fig.~\ref{fig:10}. 

\item 
We note that there are more white spaces in the middle panel of Fig.~\ref{fig:mt_analysis} than that in Fig.~\ref{fig:mm_analysis}. Because of the overall $\sin\lambda L/2$ dependence, the first term of $\delta(\Delta P^{CP}_{\mu\tau})$ (\textcolor{darkgreen}{dark green} curves in Fig.~\ref{fig:mt_analysis} and  Eq.~(\ref{eq:delta_mt}))  vanishes if $\lambda L/2 \sim \pi$ or ${L}/{E} \sim 1000$ km/GeV.  
No such overall $\sin\lambda L/2$ is present in the first term of Eq.~\ref{eq:delta_mm} for 
$\delta(\Delta P_{\mu\mu}^{CP})$, making it less probable to vanish at any value.

\item The much  smaller dark patches at energies $\lesssim 2$ GeV in the middle panels of 
Figs.~\ref{fig:9} and \ref{fig:10}  arise because of rapid oscillation
at lower energies ($\lesssim 2$ GeV) in Fig.~\ref{fig:mt_analysis} and \ref{fig:mm_analysis} respectively.

\item In presence of SI only (black dashed curves in Figs.~\ref{fig:mt_analysis} and \ref{fig:mm_analysis}), we note that the values of $\delta(\Delta P^{CP}_{\mu\tau})$ and $\delta(\Delta P^{CP}_{\mu\mu})$  are very small. This explains the almost completely white/light yellowish oscillograms in Figs.~\ref{fig:9} and \ref{fig:10} respectively (middle row, left column).
 \end{itemize}

\subsection*{$\mu-e$ sector:}
\begin{figure}[htb]
\centering
\includegraphics[width=\textwidth]
{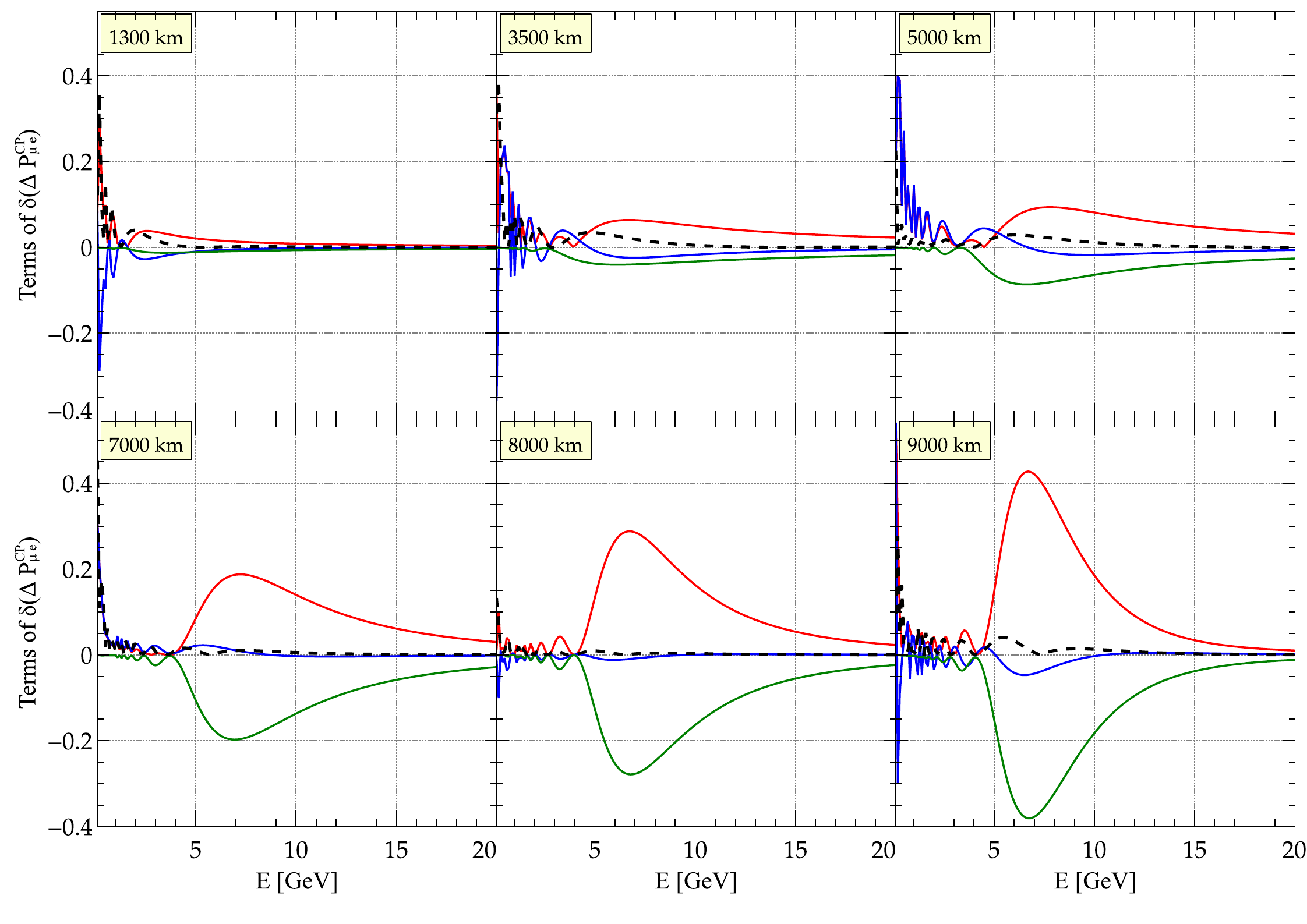}
\caption{\footnotesize{$\delta(\Delta P^{CP}_{\mu e})$ as a function of E[GeV] for 
6 fixed values of the baseline L[km]. The \textcolor{darkgreen}{darkgreen} (\textcolor{blue}{blue}) curve corresponds to the \textcolor{darkgreen}{first} (\textcolor{blue}{second} term of Eq.~(\ref{eq:delta_me})). The \textcolor{red}{red} curve is the value of $|\delta \Delta P^{CP}_{\mu e}|$ in Eq.~(\ref{eq:delta_me}). The black dashed curve corresponds to the value of $|\delta \Delta P^{CP}_{\mu e}|$ in the SI case.
 }}
\label{fig:me_analysis}
\end{figure}

\bea\label{eq:delta_me}
\delta \Delta P^{CP}_{\mu e} &=& \delta(P_{\mu e} - \bar{P}_{\mu e}) \nonumber\\
&=& (P_{\mu e} - \bar{P}_{\mu e})|_{\delta_{13} = \pi/2} - (P_{\mu e} - \bar{P}_{\mu e})|_{\delta_{13} = 0} \nonumber\\
& \approx & -2\sqrt{2} C s_{13} \bigg[ \frac{\sin^{2}((1-\hat{r_A})\lambda L/2)}{(1-\hat{r_A})^{2}} + 
\frac{\sin^{2}((1+\hat{r_A})\lambda L/2)}{(1+\hat{r_A})^{2}} \bigg] \nonumber \\
& + & \frac{2 r_\lambda \sin2\theta_{12} \sin(\hat{r_A}\lambda L/2)}{\hat{r_A}} 
\bigg[ 
CD_{1}\cos\lambda L/2 - s_{13}D_{2}\sin\lambda L/2
\nonumber \\  &-& 
\frac{\sin((1-\hat{r_A})\lambda L/2)}{1-\hat{r_A}} 
 (C + s_{13})\cos(\omega - \lambda L/2)
+ 
\frac{\sin((1+\hat{r_A})\lambda L/2)}{1+\hat{r_A}} 
(s_{13} - C)\cos(\omega + \lambda L/2)
 \bigg] \nonumber \\
&&
\eea

We make the following observations from Fig.~\ref{fig:me_analysis} which are useful to understand the features in Fig.~\ref{fig:9} (top row, middle panel).
\begin{itemize}
\item In Fig.~\ref{fig:me_analysis}, we plot $\delta(\Delta P^{CP}_{\mu e})$ as a function of $E$ for different baselines. The overall behaviour is dominated by the \textcolor{darkgreen}{first} term of Eq.~(\ref{eq:delta_me}).

 \item Unlike the case of $\delta(\Delta P^{CP}_{\mu\tau})$ or $\delta(\Delta P^{CP}_{\mu\mu})$, here we have only one primary peak in Fig.~\ref{fig:me_analysis}. This peak starts appearing roughly at $L \gtrsim 5000 km$  and rapidly grows with the baseline. This gives rise to the dark inverted triangular shaped patch in Fig.~\ref{fig:9} (top row, middle column) at $E \approx 4-6 ~GeV$. 
 
 \item The orangish blob in the same panel of Fig.~\ref{fig:9} (at $L \lesssim 2000$ km and roughly at $5~ GeV < E < 9 ~GeV$) is due to the presence of $\eee$.
 
 \end{itemize}
 
 \subsection{Pattern of CP and T oscillograms in the Sterile case}
\label{app_c}
 \begin{figure}[htb]
\centering
\includegraphics[width=\textwidth]
{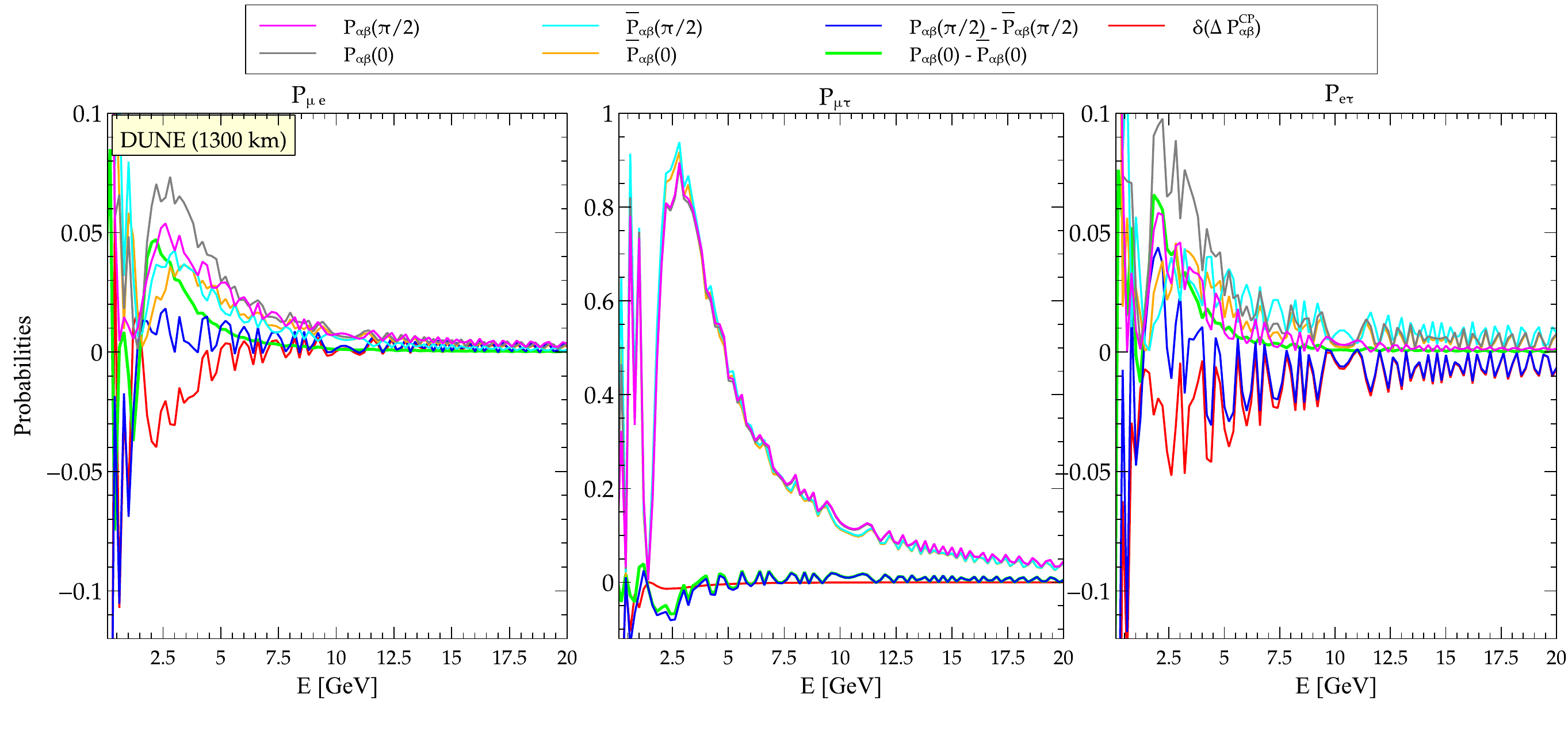}
\caption{\footnotesize{
Probability differences for the appearance channels in sterile case and size of the wiggles for different channels for a fixed baseline of 1300 km.
 }}
\label{fig:wiggle}
\end{figure}
 In Fig.~\ref{fig:wiggle}, we plot the various probability differences (see the legend) 
 that go in the calculation of $\delta \Delta P^{CP}_{\alpha\beta}$ in presence of a sterile neutrino  for the channels 
 $\nu_{\mu} \to \nu_{e}$, $\nu_{\mu} \to \nu_{\tau}$ and $\nu_{e} \to \nu_{\tau}$ 
 (corresponding to the three rows of Fig.~\ref{fig:9}). The baseline is taken to be 1300 km. 
 These three panels correspond to the three rows of the right column of Fig.~\ref{fig:9}. 
 One can see that $\delta \Delta P^{CP}_{\mu\tau}$ (represented by the red curve in Fig.~\ref{fig:wiggle}) is quite smooth, unlike $\delta \Delta P^{CP}_{\mu e}$ or $\delta \Delta P^{CP}_{e\tau}$, 
 that show rapidly oscillating nature\footnote{these rapid secondary oscillations are the manifestations of a high $\Delta m^{2}_{41} \sim 1 ~eV^{2}$}. 
 We also note that the amplitude of the wiggles is larger in the $\nu_{e} \to \nu_{\tau}$ 
 channel than that in $\nu_{\mu} \to \nu_{e}$ channel.
 Indeed, in Fig.~\ref{fig:9} (the three rows in the right column), we see that the oscillogram is mostly \textit{smooth} in the 
 $\nu_{\mu} \to \nu_{\tau}$ channel. Also, the $\nu_e \to \nu_\tau$ channel seems to be more \textit{wiggly}  
 than the $\nu_{\mu} \to \nu_{e}$ channel in the oscillogram~\footnote{Note from Table.~\ref{tab:parameters_sterile} that, we have considered a value of $15^{o}$ for $\theta_{34}$, which is quite large compared to $\theta_{14}$ ($\sim 8^{o}$) and $\theta_{24}$ ($\sim 5^{o}$). The large allowed range for $\theta_{34}$ ($< 25^{o}$) permits us to use such a large value for it. Although $\theta_{34}$ has marginal effect on the $\nu_{\mu} \to \nu_{e}$ 
 channel, the $\nu_{e} \to \nu_{\tau}$ channel depends quite significantly on $\theta_{34}$. 
 This, in turn, produces the large wiggles for $\delta \Delta P^{CP}_{e\tau}$.}.

\end{appendix}

\bibliographystyle{apsrev}
\bibliography{referencesnsi}


\end{document}